\pdfoutput=1
\RequirePackage{ifpdf}
\ifpdf 
\documentclass[pdftex]{sigma}
\else
\documentclass{sigma}
\fi

\numberwithin{equation}{section}

\catcode`\,12 
\newcommand*\pFqskip{8mu} \catcode`,\active \newcommand*\pFq{\begingroup \catcode`\,\active
\def,{\mskip\pFqskip\relax}\dopFq} \catcode`\,12
\def\dopFq#1#2#3#4#5{{}_{#1}F_{#2}\biggl[\genfrac..{0pt}{}{#3}{#4}\Big\rvert#5\biggr]\endgroup}

\newcommand{\ket}[1]{|#1\rangle}

\newcommand{\kket}[2]{|#1\rangle_{#2}}

\newcommand{\bra}[1]{\langle #1|}

\newcommand{\bbra}[2]{{}_{#2}\langle #1|}

\newcommand{\BBraket}[5]{{}_{#1}\bra{#2}#3\ket{#4}_{#5}}

\newcommand{\braket}[2]{\langle #1|#2\rangle}

\DeclareMathOperator{\tg}{tg}

\begin{document}

\newcommand{\arXivNumber}{1404.0876}

\allowdisplaybreaks

\renewcommand{\PaperNumber}{108}

\FirstPageHeading

\ShortArticleName{The Generic Superintegrable System on the 3-Sphere and the $9j$ Symbols of $\mathfrak{su}(1,1)$}
\ArticleName{The Generic Superintegrable System on the 3-Sphere\\
and the $\boldsymbol{9j}$ Symbols of $\boldsymbol{\mathfrak{su}(1,1)}$}

\Author{Vincent X.~GENEST and Luc VINET}
\AuthorNameForHeading{V.X.~Genest and L.~Vinet}

\Address{Centre de Recherches Math\'ematiques, Universit\'e de Montr\'eal,\\
P.O.~Box 6128, Centre-ville Station, Montr\'eal, QC, Canada, H3C 3J7}
\Email{\href{mailto:vincent.genest@umontreal.ca}{vincent.genest@umontreal.ca}, \href{mailto:luc.vinet@umontreal.ca}{luc.vinet@umontreal.ca}}

\ArticleDates{Received August 15, 2014, in f\/inal form November 25, 2014; Published online December 05, 2014}

\Abstract{The $9j$ symbols of $\mathfrak{su}(1,1)$ are studied within the framework of the generic superintegrable
system on the 3-sphere.
The canonical bases corresponding to the binary coupling schemes of four $\mathfrak{su}(1,1)$ representations are
constructed explicitly in terms of Jacobi polynomials and are seen to correspond to the separation of variables in
dif\/ferent cylindrical coordinate systems.
A~triple integral expression for the $9j$ coef\/f\/icients exhibiting their symmetries is derived.
A~double integral formula is obtained by extending the model to the complex three-sphere and taking the complex radius
to zero.
The explicit expression for the vacuum coef\/f\/icients is given.
Raising and lowering operators are constructed and are used to recover the relations between contiguous coef\/f\/icients.
It is seen that the $9j$ symbols can be expressed as the product of the vacuum coef\/f\/icients and a~rational function.
The recurrence relations and the dif\/ference equations satisf\/ied by the $9j$ coef\/f\/icients are derived.}


\Classification{33C50; 81R05}

\section{Introduction}

The objective of this paper is to show how the framework provided by the generic superintegrable system on the 3-sphere
can be used to study the $9j$ coef\/f\/icients of $\mathfrak{su}(1,1)$.
In addition to providing a~new interpretation for these coef\/f\/icients, this approach, which can be viewed as a~treatment
of the problem in the position representation, allows for an explicit construction of the canonical bases involved in
the~$9j$ problem and a~direct derivation of the properties of the~$9j$ symbols without reference to Clebsch--Gordan or
Racah coef\/f\/icients.

The $9j$ symbols arise as recoupling coef\/f\/icients in the combination of four irreducible $\mathfrak{su}(1,1)$
representations of the positive-discrete series.
These coef\/f\/icients and their equivalent $\mathfrak{su}(2)$ analogues have traditionally found applications in
molecular~\cite{Z-2007} and nuclear~\cite{Nuc-2007} physics but have also appeared in the study of spin networks related
to quantum gravity~\cite{Regge-2000}.
Over the past years, they have been the object of a~number of publications, many of which study the $9j$ coef\/f\/icients
from the point of view of special functions.
For example, the connection between $9j$ coef\/f\/icients and orthogonal polynomials in two variables has been studied~by
Van der Jeugt~\cite{VDJ-02-2000}, Suslov~\cite{Suslov-1991} and more recently by Hoare and Rahman~\cite{Hoare-2008} who
used the $9j$ coef\/f\/icients as a~starting point to their study of bivariate Krawtchouk
polynomials~\cite{Diaconis-02-2014, Genest-2013-06}.
A~number of explicit multi-sums expressions have also been investigated by Ali\v{s}auskas and Jucys~\cite{Alisauskas-2000, Alisauskas-1971}, Rosengren~\cite{ Rosengren-12-1998,
Rosengren-1999, Rosengren-1998} and Srinivasa~Rao and Rajeswari~\cite{Rao-1989}.
Also worth mentioning is the original approach of Granovskii and Zhedanov~\cite{Zhedanov-09-1993} which opened a~path to
a~new method for deriving generating functions and convolution identities for orthogonal polynomials~\cite{Koelink-1998, VDJ-1997}.

In the present paper, we shall indicate how the $9j$ problem can be studied in the position representation using the
connection between the coupling of four $\mathfrak{su}(1,1)$ representations and the generic superintegrable system on
the three-sphere.
This system has been discussed by Kalnins, Kress and Miller in~\cite{Kalnins-2006}.
It is governed by the Hamiltonian
\begin{gather}
\label{Hamiltonian}
H=\sum\limits_{1\leqslant i<k\leqslant4}J_{ik}^2+r^2\sum\limits_{1\leqslant \ell\leqslant 4}\frac{a_{\ell}}{s_{\ell}^2},
\qquad
a_{\ell}=\alpha_{\ell}^2-1/4,
\end{gather}
where ${\alpha}_{\ell}>-1$ and is def\/ined on the 3-sphere of square radius $r^2=s_1^2+s_2^2+s_3^2+s_4^2$.
Here the operators $J_{ik}$ stand for the familiar angular momentum generators
\begin{gather}
\label{Angular-Momentum}
J_{ik}=i(s_i\partial_{s_k}-s_k \partial_{s_i}),
\qquad
1\leqslant i<k\leqslant4.
\end{gather}
The system described by~\eqref{Hamiltonian} is both superintegrable and exactly solvable.
It has f\/ive algebraically independent second order constants of motion that generate a~quadratic
algebra~\cite{Kalnins-2011-05}.

It will f\/irst be shown that the Hamiltonian~\eqref{Hamiltonian} coincides with the total Casimir operator for the
combination of four $\mathfrak{su}(1,1)$ representations and that its constants of motion correspond to the intermediate
Casimir operators associated to each possible pairing of the four representations; these results extend the author's
previous work~\cite{Genest-2013-tmp-1}.
Using this framework, the canonical orthonormal bases of the $9j$ problem, which correspond to the joint diagonalization
of dif\/ferent pairs of commuting intermediate Casimir operators, will be constructed as solutions of the Schr\"odinger
equation associated to~\eqref{Hamiltonian} separated in dif\/ferent cylindrical coordinate systems; these solutions will
be given in terms of Jacobi polynomials.
The coordinate realization of the canonical bases and the underlying quantum mechanical framework will yield an
expression for the $9j$ coef\/f\/icients in terms of an integral on the 3-sphere exhibiting their symmetries.
By extending the model to the complex 3-sphere and taking the complex radius to zero, the expression for the $9j$
symbols in terms of a~double integral found by Granovskii and Zhedanov~\cite{Zhedanov-09-1993} shall be recovered.
This formula will be used to obtain an explicit hypergeometric formula for the special case corresponding to the
``vacuum'' $9j$ coef\/f\/icients.
The coordinate realization will also allow for the construction of raising and lowering operators based on the structure
relations of the Jacobi polynomials.
These operators will then be used to derive directly the relations between contiguous $9j$ symbols, which are usually
obtained by manipulations of Clebsch--Gordan or Racah coef\/f\/icients (given in terms of the Hahn or the Racah
polynomials~\cite{VDJ-02-2000}).
From these relations, it will be possible to conclude that the $9j$ coef\/f\/icients can be expressed as a~product of the
vacuum coef\/f\/icients and of functions that are rational (and not polynomial as stated in~\cite{Hoare-2008}).
The fact that the raising and lowering operators factorize the corresponding intermediate Casimir operators shall be
used to obtain the action of the intermediate Casimirs on the basis states.
This will also lead to both the dif\/ference equations and the recurrence relations satisf\/ied by the $9j$ coef\/f\/icients.

The organization of the paper is as follows.
\begin{itemize}
\itemsep=0pt
\item Section~\ref{Section2}: generic system on the 3-sphere from four $\mathfrak{su}(1,1)$ representations, exact solutions,
canonical basis vectors of the $9j$ problem, triple integral representation, symmetries of the $9j$ coef\/f\/icients.
\item Section~\ref{Section3}: double integral formula, explicit vacuum $9j$ coef\/f\/icients.
\item Section~\ref{Section4}: raising/lowering operators, relations between contiguous $9j$ symbols.
\item Section~\ref{Section5}: dif\/ference equations and recurrence relations for $9j$ symbols.
\end{itemize}

\section[The $9j$ problem for $\mathfrak{su}(1,1)$ in the position representation]{The $\boldsymbol{9j}$ problem for $\boldsymbol{\mathfrak{su}(1,1)}$ in the position representation}\label{Section2}

In this section the $9j$ problem for the positive-discrete series of $\mathfrak{su}(1,1)$ representations is exa\-mined in the position
representation.
The total Casimir operator for the addition of four representations is identif\/ied with the Hamiltonian of the generic
superintegrable system on~$S^3$ and the intermediate Casimir operators are identif\/ied with its symmetries.
The canonical basis vectors of the~$9j$ problem are constructed as wavefunctions separated in dif\/ferent coordinate
systems and the~$9j$ coef\/f\/icients are expressed as the overlap coef\/f\/icients between these bases.

\subsection[The addition of four representations and the generic system on $S^3$]{The addition of four representations and the generic system on $\boldsymbol{S^3}$}

Consider the operators
\begin{gather}
\label{Coordinate-Realization}
K_0^{(i)}=\frac{1}{4}\left(-\partial_{s_i}^2+s_i^2+\frac{a_i}{s_i^2}\right),
\qquad
K_{\pm}^{(i)}=\frac{1}{4}\left((s_i\mp \partial_{s_i})^2-\frac{a_i}{s_i^2}\right),
\qquad
i=1,\ldots,4,
\end{gather}
which form four mutually commuting sets of generators satisfying the $\mathfrak{su}(1,1)$ commutation relations
\begin{gather*}
\big[K_0^{(i)},K_{\pm}^{(i)}\big]=\pm K_{\pm}^{(i)},
\qquad
\big[K_{-}^{(i)},K_{+}^{(i)}\big]=2K_0^{(i)}.
\end{gather*}
The operators~\eqref{Coordinate-Realization} provide a~realization of the positive-discrete series of
$\mathfrak{su}(1,1)$ representations on the space of square-integrable functions on the positive real line.
A~set of basis vectors $e_{n_i}^{(\nu_i)}$, $n_i=0,1,\ldots$, for these representations specif\/ied by a~positive real
number $\nu_i$ taking the value
\begin{gather}
\label{Value}
\nu_i=\frac{\alpha_i+1}{2},
\end{gather}
is given in terms of Laguerre polynomials~\cite{Koekoek-2010} according to
\begin{gather}
\label{Basis-Vectors}
e_{n_i}^{(\nu_i)}(s_i)=(-1)^{n_i}\sqrt{\frac{2\Gamma(n_i+1)}{\Gamma(n_i+\alpha_i+1)}}
e^{-s_i^2/2}s_i^{\alpha_i+1/2}L_{n_i}^{(\alpha_i)}\big(s_i^2\big),
\qquad
n_i=0,1,\ldots,
\end{gather}
where $\Gamma(z)$ is the gamma function~\cite{Arfken-2012}.
These basis vectors are orthonormal with respect to the scalar product~\cite{Koekoek-2010}
\begin{gather*}
\int_{0}^{\infty}e_{n_i}^{(\nu_i)}(s_i)
e_{n_i'}^{(\nu_i)}(s_i)
\,\mathrm{d}s_i=\delta_{n_in_i'},
\end{gather*}
and the action of the generators on the basis vectors is given~by
\begin{gather*}
K_{+}^{(i)}e_{n_i}^{(\nu_i)}(s_i)=\sqrt{(n_i+1)(n_i+2\nu_i)}
e_{n_i+1}^{(\nu_i)}(s_i),
\\
K_{-}^{(i)}e_{n_i}^{(\nu_i)}(s_i)=\sqrt{n_i(n_i+2\nu_i-1)}
e_{n_i-1}^{(\nu_i)}(s_i),
\\
K_0^{(i)}e_{n_i}^{(\nu_i)}(s_i)=(n_i+\nu_i)e_{n_i}^{(\nu_i)}(s_i),
\end{gather*}
which corresponds to the usual action def\/ining the irreducible representations of the positive-discrete
series~\cite{Vilenkin-1991}.
In the realization~\eqref{Coordinate-Realization}, it is easily verif\/ied that the Casimir operator of~$\mathfrak{su}(1,1)$ which has the expression
\begin{gather*}
Q^{(i)}=\big[K_0^{(i)}\big]^2-K_{+}^{(i)}K_{-}^{(i)}-K_0^{(i)},
\end{gather*}
acts as a~multiple of the identity, i.e.
\begin{gather*}
Q^{(i)}=\nu_i(\nu_i-1),
\end{gather*}
for $i=1,\ldots,4$.
The four sets~\eqref{Coordinate-Realization} can be used to def\/ine a~f\/ifth set of $\mathfrak{su}(1,1)$ generators through
\begin{gather*}
K_0=\sum\limits_{1\leqslant i \leqslant 4}K_0^{(i)},
\qquad
K_{\pm}=\sum\limits_{1 \leqslant i\leqslant4}K_{\pm}^{(i)}.
\end{gather*}
The above operators realize a~representation of $\mathfrak{su}(1,1)$ on the space $\bigotimes_{i=1}^{4}V^{(\nu_i)}$,
where $V^{(\nu_i)}$ is the space spanned by the basis vectors~\eqref{Basis-Vectors}.
It is directly checked that the total Casimir operator associated to this realization is
\begin{gather}
\label{Total-Casimir}
Q=H/4,
\end{gather}
where~$H$ is the Hamiltonian of the generic superintegrable system on the 3-sphere given by~\eqref{Hamiltonian}.

When considering the tensor product of several representations, it is natural to consider the intermediate Casimir
operators associated to each possible pairing of representations.
These intermediate Casimir operators are def\/ined~by
\begin{gather*}
Q^{(ij)}=\big[K_0^{(i)}+K_0^{(j)}\big]^2-\big[K_{+}^{(i)}+K_{+}^{(j)}\big]\big[K_{-}^{(i)}+K_{-}^{(j)}\big]-\big[K_0^{(i)}+K_0^{(j)}\big],
\qquad
1 \leqslant i<j\leqslant 4,
\end{gather*}
and have the expression
\begin{gather}
\label{Intermediate-Casimirs}
Q^{(ij)}=\frac{1}{4}\left(J_{ij}^2+\frac{a_i s_j^2}{s_i^2}+\frac{a_j s_i^2}{s_j^2}+a_i+a_j-1\right),
\qquad
1 \leqslant i<j\leqslant 4,
\end{gather}
where $J_{ij}$ are the angular momentum operators~\eqref{Angular-Momentum}.
By construction, the intermediate Casimir operators $Q^{(ij)}$ commute with the total Casimir operator~$Q$ and hence the
intermediate Casimir operators~\eqref{Intermediate-Casimirs} are the symmetries of the Hamiltonian~\eqref{Hamiltonian}.
It is directly checked that the intermediate Casimir operators $Q^{(ij)}$, $Q^{(k\ell)}$ commute only when $i$, $j$, $k$, $\ell$
are all dif\/ferent and hence the largest set of commuting intermediate Casimir operators has two elements.
Note that the intermediate Casimir operators are linearly related to the total Casimir operator as per the relation
\begin{gather*}
Q=\sum\limits_{1 \leqslant i<j\leqslant 4}Q^{(ij)}-2\sum\limits_{1\leqslant i\leqslant 4}Q^{(i)}.
\end{gather*}
In considering the total Casimir operator~\eqref{Total-Casimir}, one can take the value of the square radius $r^2$ to be
f\/ixed since the operator
\begin{gather*}
2 K_0+K_{+}+K_{-}=r^2,
\end{gather*}
commutes with~$Q$ and all the intermediate Casimir operators $Q^{(ij)}$.
We shall take $r^2=1$, thus considering the Hamiltonian~\eqref{Hamiltonian} on the unit 3-sphere.

\subsection[The $9j$ symbols]{The $\boldsymbol{9j}$ symbols}

In general, the representation $\bigotimes_{i=1}^{4} V^{(\nu_i)}$ is not irreducible, but it is known to
be completely reducible in representations of the positive-discrete series.
In this context, the $9j$ symbols arise as the overlap coef\/f\/icients between natural bases associated to two dif\/ferent
decomposition schemes.
\begin{itemize}\itemsep=0pt
\item In the f\/irst scheme, one f\/irst decomposes $V^{(\nu_1)}\otimes V^{(\nu_2)}$ and $V^{(\nu_3)}\otimes V^{(\nu_4)}$ in
irreducible components $V^{(\nu_{12})}$, $V^{(\nu_{34})}$ and then decomposes $V^{(\nu_{12})}\otimes V^{(\nu_{34})}$ in
irreducible compo\-nents~$V^{(\nu)}$ for each occurring values of $(\nu_{12},\nu_{34})$.
The natural (orthonormal) basis vectors for this scheme are denoted $\ket{\vec{\nu};\nu_{12},\nu_{34};\nu}$ and def\/ined~by
\begin{gather}
Q^{(12)}\ket{\vec{\nu};\nu_{12},\nu_{34};\nu} =\nu_{12}(\nu_{12}-1)\ket{\vec{\nu};\nu_{12},\nu_{34};\nu},
\nonumber
\\
Q^{(34)}\ket{\vec{\nu};\nu_{12},\nu_{34};\nu} =\nu_{34}(\nu_{34}-1)\ket{\vec{\nu};\nu_{12},\nu_{34};\nu},
\nonumber
\\
Q\ket{\vec{\nu};\nu_{12},\nu_{34};\nu} =\nu(\nu-1)\ket{\vec{\nu};\nu_{12},\nu_{34};\nu},
\label{First-Scheme}
\end{gather}
where $\vec{\nu}=(\nu_1,\nu_2,\nu_3,\nu_4)$.
\item In the second scheme, one f\/irst decomposes $V^{(\nu_1)}\otimes V^{(\nu_3)}$ and $V^{(\nu_2)}\otimes V^{(\nu_4)}$
in irreducible components $V^{(\nu_{13})}$, $V^{(\nu_{24})}$ and then decomposes $V^{(\nu_{13})}\otimes V^{(\nu_{24})}$
in irreducible compo\-nents~$V^{(\nu)}$ for each occurring values of $(\nu_{13},\nu_{24})$.
The natural (orthonormal) basis vectors for this scheme are denoted $\ket{\vec{\nu};\nu_{13},\nu_{24};\nu}$ and def\/ined~by
\begin{gather}
Q^{(13)}\ket{\vec{\nu};\nu_{13},\nu_{24};\nu} =\nu_{13}(\nu_{13}-1)\ket{\vec{\nu};\nu_{13},\nu_{24};\nu},
\nonumber
\\
Q^{(24)}\ket{\vec{\nu};\nu_{13},\nu_{24};\nu} =\nu_{24}(\nu_{24}-1)\ket{\vec{\nu};\nu_{13},\nu_{24};\nu},
\nonumber
\\
Q\ket{\vec{\nu};\nu_{13},\nu_{24};\nu} =\nu(\nu-1)\ket{\vec{\nu};\nu_{13},\nu_{24};\nu}.
\label{Second-Scheme}
\end{gather}
\end{itemize}
The $9j$ symbols are def\/ined as the overlap coef\/f\/icients between these two bases, i.e.
\begin{gather*}
\ket{\vec{\nu};\nu_{12},\nu_{34};\nu}=\sum\limits_{\nu_{13},\nu_{24}}
\begin{Bmatrix}
\nu_1	 & \nu_2	 & \nu_{12}
\\
\nu_3	 & \nu_4	 & \nu_{34}
\\
\nu_{13} & \nu_{24} & \nu
\end{Bmatrix}
\ket{\vec{\nu};\nu_{13},\nu_{24};\nu}.
\end{gather*}
For the $9j$ symbols to be non-vanishing, one must have
\begin{gather}
\nu_{12}=\nu_1+\nu_2+m,
\qquad
\nu_{34}=\nu_3+\nu_4+n,
\qquad
\nu_{13}=\nu_1+\nu_3+x,
\qquad
\nu_{24}=\nu_2+\nu_4+y,
\nonumber
\\
\nu=\nu_1+\nu_2+\nu_3+\nu_4+N,
\label{Conditions}
\end{gather}
where $m$, $n$, $x$, $y$ and~$N$ are non-negative integers such that $m+n\leqslant N$ and $x+y\leqslant N$.

In view of the coordinate realization stemming from the previous subsection, the bases~\eqref{First-Scheme}
and~\eqref{Second-Scheme} can be constructed explicitly by solving the corresponding eigenvalue equations: these bases
correspond to the diagonalization of the Hamiltonian~\eqref{Hamiltonian} together with the pairs of commuting
intermediate Casimir operators (symmetries) $(Q^{(12)}, Q^{(34)})$ or $(Q^{(13)},Q^{(24)})$.
In view of the conditions~\eqref{Value},~\eqref{Conditions} and for notational convenience, the basis corresponding to
the scheme~\eqref{First-Scheme} shall be simply denoted by $\ket{m,n}_{N}$, the basis corresponding
to~\eqref{Second-Scheme} by $\ket{x,y}_{N}$ and the $9j$ coef\/f\/icients will be written as
\begin{gather}
\label{9j-Def}
\ket{m,n}_{N}=\sum\limits_{\substack{x,y
\\
x+y\leqslant N}}
\begin{Bmatrix}
\alpha_1 & \alpha_2 & m
\\
\alpha_3 & \alpha_4 & n
\\
x & y & N
\end{Bmatrix}
\ket{x,y}_{N},
\end{gather}
or equivalently as
\begin{gather*}
\begin{Bmatrix}
\alpha_1 & \alpha_2 & m
\\
\alpha_3 & \alpha_4 & n
\\
x & y & N
\end{Bmatrix}
={}_{N}\braket{x,y}{m,n}_{N}.
\end{gather*}
The $9j$ coef\/f\/icients are taken to be real.
Since they are transition coef\/f\/icients between two orthonormal bases, it follows from elementary linear algebra that
\begin{gather}
\label{9j-Ortho}
\sum\limits_{\substack{x,y
\\
x+y\leqslant N}}
\begin{Bmatrix}
\alpha_1 & \alpha_2 & m
\\
\alpha_3 & \alpha_4 & n
\\
x & y & N
\end{Bmatrix}
\begin{Bmatrix}
\alpha_1 & \alpha_2 & m'
\\
\alpha_3 & \alpha_4 & n'
\\
x & y & N
\end{Bmatrix}
=\delta_{mm'}\delta_{nn'},
\end{gather}
and similarly
\begin{gather*}
\sum\limits_{\substack{m,n
\\
m+n\leqslant N}}
\begin{Bmatrix}
\alpha_1 & \alpha_2 & m
\\
\alpha_3 & \alpha_4 & n
\\
x & y & N
\end{Bmatrix}
\begin{Bmatrix}
\alpha_1 & \alpha_2 & m
\\
\alpha_3 & \alpha_4 & n
\\
x' & y' & N
\end{Bmatrix}
=\delta_{xx'}\delta_{yy'}.
\end{gather*}

\subsection{The canonical bases by separation of variables}

Let us now obtain the explicit realizations for the bases $\ket{m,n}_{N}$ and $\ket{x,y}_{N}$ corresponding to the
coupling schemes~\eqref{First-Scheme} and~\eqref{Second-Scheme}.
As shall be seen, these bases correspond to the separation of variables in the equation $H\Upsilon=\Lambda \Upsilon$
using dif\/ferent cylindrical coordinate systems.
Note that this eigenvalue equation has been studied by Kalnins, Miller and Tratnik in~\cite{Kalnins-1991}.

\subsubsection[The basis for $\{Q^{(12)},Q^{(34)}\}$]{The basis for $\boldsymbol{\{Q^{(12)},Q^{(34)}\}}$}

To obtain the coordinate realization of the basis corresponding to the f\/irst coupling
scheme~\eqref{First-Scheme}, we look for functions $\Psi_{m,n;N}$ on the 3-sphere that satisfy
\begin{gather*}
Q^{(12)}\Psi_{m,n;N}=\lambda^{(12)}_{m}\Psi_{m,n;N},
\qquad
Q^{(34)}\Psi_{m,n;N}=\lambda^{(34)}_{n}\Psi_{m,n;N},
\qquad
Q\Psi_{m,n;N}=\Lambda_{N}\Psi_{m,n;N},
\end{gather*}
with eigenvalues
\begin{gather*}
\lambda^{(12)}_{m} =(m+\alpha_1/2+\alpha_2/2)(m+\alpha_1/2+\alpha_2/2+1),
\\
\lambda^{(34)}_{n} =(n+\alpha_3/2+\alpha_4/2)(n+\alpha_3/2+\alpha_4/2+1),
\\
\Lambda_{N} =(N+|\alpha|/2+1)(N+|\alpha|/2+2),
\end{gather*}
where $|\alpha|=\sum\limits_{i}\alpha_i$.
The expressions for the spectra follow directly from the fact that the operators are intermediate Casimir operators in
the addition of $\mathfrak{su}(1,1)$ representations of the positive-discrete series~\cite{VDJ-2003}.
Consider the set of cylindrical coordinates $\{\theta,\phi_1,\phi_2\}$ def\/ined~by
\begin{gather}
\label{Coords-1}
s_1=\cos\theta \cos \phi_1,
\qquad
s_2=\cos\theta \sin\phi_1,
\qquad
s_3=\sin\theta \cos \phi_2,
\qquad
s_4=\sin\theta \sin \phi_2.
\end{gather}
In these coordinates, one f\/inds from~\eqref{Intermediate-Casimirs} that the operators $Q^{(12)}$, $Q^{(34)}$ read
\begin{gather*}
Q^{(12)} =\frac{1}{4}\left(-\partial_{\phi_1}^2+a_1
\tg^2\phi_1+\frac{a_2}{\tg^2\phi_1}+(a_1+a_2-1)\right),
\\
Q^{(34)} =\frac{1}{4}\left(-\partial_{\phi_2}^2+a_3
\tg^2\phi_2+\frac{a_4}{\tg^2\phi_2}+(a_3+a_4-1)\right),
\end{gather*}
and that~$Q$ takes the form
\begin{gather*}
Q=\frac{1}{4}\bigg[-\partial_{\theta}^2+\left(\tg\theta-\frac{1}{\tg\theta}\right)\partial_{\theta}
\\
\phantom{Q=}{}
+\frac{1}{\cos^2\theta}\left(-\partial_{\phi_1}^2+\frac{a_1}{\cos^2\phi_1}+\frac{a_2}{\sin^2\phi_1}\right)
+\frac{1}{\sin^2\theta}\left(-\partial_{\phi_2}^2+\frac{a_3}{\cos^2\phi_2}+\frac{a_4}{\sin^2\phi_2}\right)\bigg].
\end{gather*}
It is directly seen from the above expressions that $\Psi_{m,n;N}$ will separate in the coordinates~\eqref{Coords-1}.
Using standard techniques, one f\/inds that the wavefunctions have the expression
\begin{gather}
\braket{\theta,\phi_1,\phi_2}{m,n}_{N}=\Psi_{m,n;N}^{(\alpha_1,\alpha_2,\alpha_3,\alpha_4)}(\theta,\phi_1,\phi_2)
=\eta_{m}^{(\alpha_2,\alpha_1)}\eta_{n}^{(\alpha_4,\alpha_3)}
\nonumber
\\
\qquad{}
\times
\eta_{N-m-n}^{(2n+\alpha_3+\alpha_4+1,2m+\alpha_1+\alpha_2+1)}  (\cos \theta \cos \phi_1)^{\alpha_1+1/2}
(\cos \theta\sin\phi_1)^{\alpha_2+1/2}
\nonumber
\\
\qquad{}
\times
(\sin\theta \cos \phi_2)^{\alpha_3+1/2}
(\sin\theta\sin\phi_2)^{\alpha_4+1/2}
\cos^{2m}\theta\sin^{2n}\theta
P_{m}^{(\alpha_2,\alpha_1)}(\cos 2\phi_1)
\nonumber
\\
\qquad{}
\times
P_{n}^{(\alpha_4,\alpha_3)}(\cos 2\phi_2)
P_{N-m-n}^{(2n+\alpha_3+\alpha_4+1,2m+\alpha_1+\alpha_2+1)}(\cos 2\theta),
\label{Wave-Scheme-1}
\end{gather}
where $P_{n}^{(\alpha,\beta)}(x)$ are the classical Jacobi polynomials (see Appendix~\ref{appendixA}).
The normalization factor
\begin{gather}
\label{Normalization}
\eta_{n}^{(\alpha,\beta)}
=\sqrt{\frac{2\,\Gamma(m+1)\Gamma(m+\alpha+\beta+1)\Gamma(2m+\alpha+\beta+2)}{\Gamma(m+\alpha+1)\Gamma(m+\beta+1)\Gamma(2m+\alpha+\beta+1)}},
\end{gather}
ensures that the following orthonormality condition holds:
\begin{gather}
\label{Ortho-1}
\int_{0}^{\pi/2}\int_{0}^{\pi/2}\int_{0}^{\pi/2}{}_{N}\braket{m',n'}{\theta,\phi_1,\phi_2}\braket{\theta,\phi_1,\phi_2}{m,n}_{N}\,\mathrm{d}\Omega
=\delta_{mm'}\delta_{nn'}\delta_{NN'},
\end{gather}
where $\mathrm{d}\Omega=\cos\theta\sin\theta
\,
\mathrm{d}\theta\,\mathrm{d}\phi_1\,\mathrm{d}\phi_2$.
In Cartesian coordinates, $\Psi_{m,n;N}^{(\alpha_1,\alpha_2,\alpha_3,\alpha_4)}$ assumes the form
\begin{gather}
\braket{s_1,s_2,s_3,s_4}{m,n}_{N}=\Psi_{m,n;N}^{(\alpha_1,\alpha_2,\alpha_3,\alpha_4)}(s_1,s_2,s_3,s_4)=
\eta_{m}^{(\alpha_2,\alpha_1)}\eta_{n}^{(\alpha_4,\alpha_3)}
\nonumber
\\
\qquad{}
\times
\eta_{N-m-n}^{(2n+\alpha_3+\alpha_4+1,2m+\alpha_1+\alpha_2+1)}
\left(\prod\limits_{i=1}^{4}s_i^{\alpha_i+1/2}\right) \big(s_1^2+s_2^2\big)^{m}\big(s_3^2+s_4^2\big)^{n}
P_{m}^{(\alpha_2,\alpha_1)}\left(\frac{s_1^2-s_2^2}{s_1^2+s_2^2}\right)
\nonumber
\\
\qquad{}
\times
P_{n}^{(\alpha_4,\alpha_3)}\left(\frac{s_3^2-s_4^2}{s_3^2+s_4^2}\right)
P_{N-m-n}^{(2n+\alpha_3+\alpha_4+1,2m+\alpha_1+\alpha_2+1)}\big(s_1^2+s_2^2-s_3^2-s_4^2\big).
\label{Psi}
\end{gather}
The wavefunctions $\Psi_{m,n;N}$ thus provide a~concrete realization in the position representation of the basis state
$\ket{m,n}_{N}$ corresponding to the f\/irst coupling scheme.
A~dif\/ferent realization of this state is given by Lievens and Van der Jeugt in~\cite{VDJ-2003-2}, who examined
realizations of coupled vectors in the coherent state representation for general tensor products.

\subsubsection[The basis for $\{Q^{(13)},Q^{(24)}\}$]{The basis for $\boldsymbol{\{Q^{(13)},Q^{(24)}\}}$}

 To obtain the coordinate realization of the basis corresponding to the second coupling
sche\-me~\eqref{Second-Scheme}, we look for functions $\Xi_{x,y;N}$ on the 3-sphere that satisfy
\begin{gather*}
Q^{(13)}\Xi_{x,y;N}=\lambda_{x}^{(13)}\Xi_{x,y;N},
\qquad
Q^{(24)}\Xi_{x,y;N}=\lambda_{y}^{(24)}\Xi_{x,y;N},
\qquad
Q\Xi_{x,y;N}=\Lambda_{N}\Xi_{x,y;N},
\end{gather*}
where
\begin{gather*}
\lambda_{x}^{(13)} =(x+\alpha_1/2+\alpha_3/2)(x+\alpha_1/2+\alpha_3/2+1),
\\
\lambda_{y}^{(24)} =(y+\alpha_2/2+\alpha_4/2)(y+\alpha_2/2+\alpha_4/2+1),
\\
\Lambda_{N} =(N+|\alpha|/2+1)(N+|\alpha|/2+2),
\end{gather*}
and $|\alpha|=\sum\limits_{i=1}^{4}\alpha_i$.
Consider the set of cylindrical coordinates $\{\vartheta,\varphi_{1},\varphi_{2}\}$ def\/ined~by
\begin{gather}
\label{Coords-2}
s_1=\cos \vartheta\cos\varphi_1,
\qquad
s_2=\sin\vartheta\cos\varphi_2,
\qquad
s_3=\cos \vartheta\sin\varphi_1,
\qquad
s_4=\sin\vartheta\sin\varphi_2.
\end{gather}
In these coordinates, the operators $Q^{(13)}$, $Q^{(24)}$ have the expressions
\begin{gather*}
Q^{(13)}=\frac{1}{4}\left(-\partial_{\varphi_1}^2+a_1 \tg^2\varphi_1+\frac{a_3}{\tg^2
\varphi_1}+(a_1+a_3-1)\right),
\\
Q^{(24)}=\frac{1}{4}\left(-\partial_{\varphi_2}^2+a_2 \tg^2
\varphi_2+\frac{a_4}{\tg^2\varphi_2}+(a_2+a_4-1)\right),
\end{gather*}
and the total Casimir operator~$Q$ reads
\begin{gather*}
Q=\frac{1}{4}\bigg[{-}\partial_{\vartheta}^2+\left(\tg
\vartheta+\frac{1}{\tg
\vartheta}\right)\partial_{\vartheta}
\\
\phantom{Q=}{}
+\frac{1}{\cos^2\vartheta}\left(-\partial_{\varphi_1}^2+\frac{a_1}{\cos^2\varphi_1}+\frac{a_3}{\sin^2\varphi_1}\right)
+\frac{1}{\sin^2\vartheta}\left(-\partial_{\varphi_2}+\frac{a_2}{\cos^2\varphi_2}+\frac{a_4}{\sin^2\varphi_2}\right)\bigg].
\end{gather*}
It is clear from the above that the functions $\Xi_{x,y;N}$ will separate in the coordinates~\eqref{Coords-2}.
The wavefunctions $\Xi_{x,y;N}$ have the expression
\begin{gather*}
\braket{\vartheta,\varphi_1,\varphi_2}{x,y}_{N}=\Xi_{x,y;N}^{(\alpha_1,\alpha_2,\alpha_3,\alpha_4)}(\vartheta,\varphi_1,\varphi_2)=
\eta_{x}^{(\alpha_3,\alpha_1)}\eta_{y}^{(\alpha_4,\alpha_2)}
\\
\qquad{}
\times
\eta_{N-x-y}^{(2y+\alpha_2+\alpha_4+1,2x+\alpha_1+\alpha_3+1)} (\cos \vartheta \cos \varphi_1)^{\alpha_1+1/2}
(\sin \vartheta\cos\varphi_2)^{\alpha_2+1/2}
\\
\qquad{}
\times
(\cos\vartheta \sin \varphi_1)^{\alpha_3+1/2}
(\sin\vartheta\sin\varphi_2)^{\alpha_4+1/2}
\cos^{2x}\vartheta\sin^{2y}\vartheta
P_{x}^{(\alpha_3,\alpha_1)}(\cos 2\varphi_1)
\\
\qquad{}
\times
P_{y}^{(\alpha_4,\alpha_2)}(\cos 2\varphi_2)
P_{N-x-y}^{(2y+\alpha_2+\alpha_4+1,2x+\alpha_1+\alpha_3+1)}(\cos 2\vartheta),
\end{gather*}
where $\eta_{n}^{(\alpha,\beta)}$ is given by~\eqref{Normalization} and where $P_{n}^{(\alpha,\beta)}(x)$ are again the
classical Jacobi polynomials.
The wavefunctions obey the orthonormality condition
\begin{gather}
\label{Ortho-2}
\int_{0}^{\pi/2}\int_{0}^{\pi/2}\int_{0}^{\pi/2}
{}_{N}\braket{x',y'}{\vartheta,\varphi_1,\varphi_2}
\braket{\vartheta,\varphi_1,\varphi_2}{x,y}_{N}
\mathrm{d}\Omega=\delta_{xx'}\delta_{yy'}\delta_{NN'},
\end{gather}
where $\mathrm{d}\Omega=\cos\vartheta\sin\vartheta\,\mathrm{d}\vartheta\,\mathrm{d}\varphi_1\,\mathrm{d}\varphi_2$.
In Cartesian coordinates, one has
\begin{gather}
\braket{s_1,s_2,s_3,s_4}{x,y}_{N}=\Xi_{x,y;N}^{(\alpha_1,\alpha_2,\alpha_3,\alpha_4)}(s_1,s_2,s_3,s_4)
=\eta_{x}^{(\alpha_3,\alpha_1)}\eta_{y}^{(\alpha_4,\alpha_2)}
\nonumber
\\
\qquad{}
\times
\eta_{N-x-y}^{(2y+\alpha_2+\alpha_4+1,2x+\alpha_1+\alpha_3+1)}
\left(\prod\limits_{i=1}^{4}s_i^{\alpha_i+1/2}\right)\big(s_1^2+s_3^2\big)^{x}\big(s_2^2+s_4^2\big)^{y}
P_{x}^{(\alpha_3,\alpha_1)}\left(\frac{s_1^2-s_3^2}{s_1^2+s_3^2}\right)
\nonumber
\\
\qquad{}
\times
P_{y}^{(\alpha_4,\alpha_2)}\left(\frac{s_2^2-s_4^2}{s_2^2+s_4^2}\right)
P_{N-x-y}^{(2y+\alpha_2+\alpha_4+1,2x+\alpha_1+\alpha_3+1)}\big(s_1^2+s_3^2-s_2^2-s_4^2\big).
\label{Xi}
\end{gather}
Note that~\eqref{Xi} can be obtained directly from~\eqref{Psi} by permuting the indices $2$ and~$3$.
The wavefunctions $\Xi_{x,y;N}$ thus provide a~concrete realization of the basis states $\ket{x,y}_{N}$ corresponding to
the coupling scheme~\eqref{Second-Scheme} in the position representation.

\subsection[$9j$ symbols as overlap coef\/f\/icients, integral representation and symmetries]{$\boldsymbol{9j}$ symbols as overlap coef\/f\/icients, integral representation\\ and symmetries}

In view of~\eqref{9j-Def}, the $9j$
coef\/f\/icients for the positive-discrete series of $\mathfrak{su}(1,1)$ representations can be expressed as the expansion
coef\/f\/icients between the wavefunctions $\Psi_{m,n;N}$ and $\Xi_{x,y;N}$ at a~given point, i.e.
\begin{gather}
\label{Expansion}
\Psi_{m,n;N}=\sum\limits_{\substack{x,y
\\
x+y\leq N}}
\begin{Bmatrix}
\alpha_1 & \alpha_2 & m
\\
\alpha_3 & \alpha_4 & n
\\
x &  y &  N
\end{Bmatrix}
\Xi_{x,y;N}.
\end{gather}
The orthogonality relation~\eqref{Ortho-2} immediately yields the integral formula
\begin{gather}
\begin{Bmatrix}
\alpha_1 & \alpha_2 & m
\\
\alpha_3 & \alpha_4 & n
\\
x &  y &  N
\end{Bmatrix}
=\eta_{m}^{(\alpha_2,\alpha_1)}\eta_{n}^{(\alpha_4,\alpha_3)}\eta_{N-m-n}^{(2n+\alpha_3+\alpha_4+1,2m+\alpha_1+\alpha_2+1)}
\nonumber
\\
\qquad{}
\times
\eta_{x}^{(\alpha_3,\alpha_1)}\eta_{y}^{(\alpha_4,\alpha_2)}\eta_{N-x-y}^{(2y+\alpha_2+\alpha_4+1,2x+\alpha_1+\alpha_3+1)}\int_{S^3_{+}}
\prod\limits_{i=1}^{4}\big(s_i^2\big)^{\alpha_i+1/2}\mathrm{d}s_i
\big(s_1^2+s_2^2\big)^{m}\big(s_3^2+s_4^2\big)^{n}
\nonumber
\\
\qquad{}
\times\big(s_1^2+s_3^2\big)^{x}\big(s_2^2+s_4^2\big)^{y}
P_{m}^{(\alpha_2,\alpha_1)}\left(\frac{s_1^2-s_2^2}{s_1^2+s_2^2}\right)
P_{n}^{(\alpha_4,\alpha_3)}\left(\frac{s_3^2-s_4^2}{s_3^2+s_4^2}\right)
P_{x}^{(\alpha_3,\alpha_1)}\left(\frac{s_1^2-s_3^2}{s_1^2+s_3^2}\right)
\nonumber
\\
\qquad{}
\times
P_{y}^{(\alpha_4,\alpha_2)}\left(\frac{s_2^2-s_4^2}{s_2^2+s_4^2}\right)
P_{N-m-n}^{(2n+\alpha_3+\alpha_4+1,2m+\alpha_1+\alpha_2+1)}\big(s_1^2+s_2^2-s_3^2-s_4^2\big)
\nonumber
\\
\qquad{}
\times
P_{N-x-y}^{(2y+\alpha_2+\alpha_4+1,2x+\alpha_1+\alpha_3+1)}\big(s_1^2+s_3^2-s_2^2-s_4^2\big),
\label{9j-Integral-1}
\end{gather}
where $S^3_{+}$ stands for the totally positive octant of the 3-sphere described by $\sum\limits_{i=1}^{4}s_i^2=1$ with
$s_i\geqslant 0$.
The integral expression~\eqref{9j-Integral-1} looks rather complicated and shall be simplif\/ied in the next section.
However, the formula~\eqref{9j-Integral-1} and the elementary properties of the Jacobi polynomials can be used to
ef\/f\/iciently obtain the symmetry relations satisf\/ied by the $9j$ symbols~\eqref{9j-Def}.
As a~f\/irst example, one can read of\/f directly from~\eqref{9j-Integral-1} the symmetry relation
\begin{gather}
\label{Duality}
\begin{Bmatrix}
\alpha_1 & \alpha_2 & m
\\
\alpha_3 & \alpha_4 & n
\\
x &  y &  N
\end{Bmatrix}
=
\begin{Bmatrix}
\alpha_1 & \alpha_3 & x
\\
\alpha_2 & \alpha_4 & y
\\
m &  n &  N
\end{Bmatrix}
.
\end{gather}
which we shall refer to as the ``duality property'' of $9j$ symbols.
As a~second example, using the well-known identity $P_{n}^{(\alpha,\beta)}(-x)=(-1)^{n}P_{n}^{(\beta,\alpha)}(x)$, one
f\/inds that
\begin{gather}
\begin{Bmatrix}
\alpha_1 & \alpha_2 & m
\\
\alpha_3 & \alpha_4 & n
\\
x &  y &  N
\end{Bmatrix}
\!=\!(-1)^{N{+}m{+}n{-}x{-}y}\!
\begin{Bmatrix}
\alpha_2 & \alpha_1 & m
\\
\alpha_4 & \alpha_3 & n
\\
y &  x &  N
\end{Bmatrix}
\!=\!(-1)^{N{+}x{+}y{-}m{-}n}\!
\begin{Bmatrix}
\alpha_3 & \alpha_4 & n
\\
\alpha_1 & \alpha_2 & m
\\
x &  y &  N
\end{Bmatrix}\!.\!\!\!\!\!
\label{Duality-2}
\end{gather}
A~number of other symmetries can be derived by combining the above.
Let us note that the formula~\eqref{9j-Integral-1} can also be found from the results of~\cite{Lievens-2002}.

\section[Double integral formula and the vacuum $9j$ coef\/f\/icients]{Double integral formula and the vacuum $\boldsymbol{9j}$ coef\/f\/icients}\label{Section3}

In this section, a~double integral formula for the $9j$ symbols is obtained by extending the
wavefunctions to the complex three-sphere and taking the complex radius to zero.
The formula is then used to compute the vacuum $9j$ coef\/f\/icients explicitly.

\subsection{Extension of the wavefunctions}

The wavefunctions $\Psi_{m,n;N}$ and $\Xi_{x,y;N}$ can easily be extended to the complex three-sphere of radius $r^2$
using their expressions in Cartesian coordinates.
The extended wavefunctions $\widetilde{\Psi}_{m,n;N}$, $\widetilde{\Xi}_{x,y;N}$ have the expressions
\begin{gather}
\widetilde{\Psi}_{m,n;N}=\eta_{m}^{(\alpha_2,\alpha_1)}\eta_{n}^{(\alpha_4,\alpha_3)}\eta_{N-m-n}^{(2n+\alpha_3+\alpha_4+1,2m+\alpha_1+\alpha_2+1)}
\left(\prod\limits_{i=1}^{4}s_i^{\alpha_i+1/2}\right)
\nonumber
\\
\phantom{\widetilde{\Psi}_{m,n;N}=}{}
\times
\big(s_1^2+s_2^2\big)^{m}\big(s_3^2+s_4^2\big)^{n}\big(s_1^2+s_2^2+s_3^2+s_4^2\big)^{N-m-n}
P_{m}^{(\alpha_2,\alpha_1)}\left(\frac{s_1^2-s_2^2}{s_1^2+s_2^2}\right)
\nonumber
\\
\phantom{\widetilde{\Psi}_{m,n;N}=}{}
\times
P_{n}^{(\alpha_4,\alpha_3)}\left(\frac{s_3^2-s_4^2}{s_3^2+s_4^2}\right)
P_{N-m-n}^{(2n+\alpha_3+\alpha_4+1,2m+\alpha_1+\alpha_2+1)}\left(\frac{s_1^2+s_2^2-s_3^2-s_4^2}{s_1^2+s_2^2+s_3^2+s_4^2}\right),
\label{Wave-1}
\end{gather}
and
\begin{gather}
\widetilde{\Xi}_{x,y;N}=\eta_{x}^{(\alpha_3,\alpha_1)}\eta_{y}^{(\alpha_4,\alpha_2)}\eta_{N-x-y}^{(2y+\alpha_2+\alpha_4+1,2x+\alpha_1+\alpha_3+1)}
\left(\prod\limits_{i=1}^{4}s_i^{\alpha_i+1/2}\right)
\nonumber
\\
\phantom{\widetilde{\Xi}_{x,y;N}=}{}
\times
\big(s_1^2+s_3^2\big)^{x}\big(s_2^2+s_4^2\big)^{y}\big(s_1^2+s_2^2+s_3^2+s_4^2\big)^{N-x-y}
P_{x}^{(\alpha_3,\alpha_1)}\left(\frac{s_1^2-s_3^2}{s_1^2+s_3^2}\right)
\nonumber
\\
\phantom{\widetilde{\Xi}_{x,y;N}=}{}
\times
P_{y}^{(\alpha_4,\alpha_2)}\left(\frac{s_2^2-s_4^2}{s_2^2+s_4^2}\right)
P_{N-x-y}^{(2y+\alpha_2+\alpha_4+1,2x+\alpha_1+\alpha_3+1)}\left(\frac{s_1^2+s_3^2-s_2^2-s_4^2}{s_1^2+s_2^2+s_3^2+s_4^2}\right),
\label{Wave-2}
\end{gather}
with $s_i\in\mathbb{C}$ for $i=1,\ldots,4$.
The expressions~\eqref{Wave-1} and~\eqref{Wave-2} correspond to the bases constructed by Lievens and Van der Jeugt
in~\cite{Lievens-2002} in their examination of $3nj$ symbols for $\mathfrak{su}(1,1)$.
The basis vectors~\eqref{Wave-1} and~\eqref{Wave-2} also resemble the harmonic functions on $S^{3}$ of Dunkl and
Xu~\cite{Dunkl-2001}, but do not correspond to the same separation of variables.

When the coordinates satisfy $s_1^2+s_2^2+s_3^2+s_4^2=1$, the wavefunctions~\eqref{Wave-1} and~\eqref{Wave-2} coincide
with~\eqref{Psi} and~\eqref{Xi}, respectively.
When $r^2\neq 1$, $\widetilde{\Psi}_{m,n;N}$ and $\widetilde{\Xi}_{x,y;N}$ dif\/fer from $\Psi_{m,n;N}$ and $\Xi_{x,y;N}$
by a~constant factor of $r^{N+|\alpha|+2}$.
Since the parameters~$N$ and $\alpha_i$ are f\/ixed, the expansion~\eqref{Expansion} is not af\/fected by this common
multiplicative factor and one can write
\begin{gather}
\label{Expansion-2}
\widetilde{\Psi}_{m,n;N}(s_1,s_2,s_3,s_4)= \sum\limits_{\substack{x,y
\\
x+y\leq N}}
\begin{Bmatrix}
\alpha_1 & \alpha_2 & m
\\
\alpha_3 & \alpha_4 & n
\\
x &  y &  N
\end{Bmatrix}
\widetilde{\Xi}_{x,y;N}(s_1,s_2,s_3,s_4),
\end{gather}
for a~given point $(s_1,s_2,s_3,s_4)$ satisfying $s_1^2+s_2^2+s_3^2+s_4^2=r^2$.
Let us now impose the condition
\begin{gather*}
s_1^2+s_2^2=-\big(s_3^2+s_4^2\big),
\end{gather*}
which corresponds to taking the radius of the complex three-sphere to zero.
Upon introducing the new variables~$u$ and~$v$ def\/ined~by
\begin{gather*}
u=\frac{s_1^2-s_3^2}{s_1^2+s_3^2},
\qquad
v=\frac{s_3^2+2s_4^2+s_1^2}{s_1^2+s_3^2},
\end{gather*}
and using the identity
\begin{gather*}
(x+y)^{m}P_{m}^{(\alpha,\beta)}\left(\frac{x-y}{x+y}\right)=\frac{(\alpha+1)_{m}}{m!}x^{m}
\pFq{2}{1}{-m,-\beta-m}{\alpha+1}{-\frac{y}{x}},
\end{gather*}
in~\eqref{Wave-1} and~\eqref{Wave-2}, one f\/inds that the expansion~\eqref{Expansion-2} reduces to
\begin{gather*}
c_{m;n;N}\left(\frac{u-v}{2}\right)^{N}P_{m}^{(\alpha_2,\alpha_1)}\left(\frac{u+v+2}{u-v}\right)\,
P_{n}^{(\alpha_4,\alpha_3)}\left(\frac{2-u-v}{v-u}\right)
\\
\qquad{}
= \sum\limits_{\substack{x,y\\x+y\leqslant N}}
\begin{Bmatrix}
\alpha_1 & \alpha_2 & m
\\
\alpha_3 & \alpha_4 & n
\\
x &  y &  N
\end{Bmatrix}
d_{x,y;N}\,P_{x}^{(\alpha_3,\alpha_1)}(u)\,P_{y}^{(\alpha_4,\alpha_2)}(v),
\end{gather*}
where the coef\/f\/icients $c_{m,n;N}$ and $d_{x,y;N}$ read
\begin{gather*}
c_{m,n;N} =\eta_{m}^{(\alpha_2,\alpha_1)}\eta_{n}^{(\alpha_4,\alpha_3)}\eta_{N-m-n}^{(2n+\alpha_3+\alpha_4+1,2m+\alpha_1+\alpha_2+1)}
\frac{(-1)^{n}(N+m+n+|\alpha|+3)_{N-m-n}}{(N-m-n)!},
\\
d_{x,y;N} =
\eta_{x}^{(\alpha_3,\alpha_1)}\eta_{y}^{(\alpha_4,\alpha_2)}\eta_{N-x-y}^{(2y+\alpha_2+\alpha_4+1,2x+\alpha_1+\alpha_3+1)}
\frac{(-1)^{y}(N+x+y+|\alpha|+3)_{N-x-y}}{(N-x-y)!}.
\end{gather*}
Here $(a)_{n}$ stands for the Pochhammer symbol def\/ined~by
\begin{gather*}
(a)_{n}=(a)(a+1)\cdots(a+n-1),
\qquad
(a)_0=1.
\end{gather*}
The orthogonality relation~\eqref{Ortho-Jacobi} for the Jacobi polynomials then leads to the integral representation
\begin{gather}
\begin{Bmatrix}
\alpha_1 & \alpha_2 & m
\\
\alpha_3 & \alpha_4 & n
\\
x &  y &  N
\end{Bmatrix}
=\left[\frac{c_{m,n;N}}{d_{x,y;N}}\frac{2^{-N}}{h_{x}^{(\alpha_3,\alpha_1)}h_{y}^{(\alpha_4,\alpha_2)}}\right]
\nonumber
\\
\qquad{}
\times \int_{-1}^{1}\int_{-1}^{1}\mathrm{d}u\,\mathrm{d}v(1-u)^{\alpha_3}(1+u)^{\alpha_1}(1-v)^{\alpha_4}(1+v)^{\alpha_2}
\nonumber
\\
\qquad{}
\times
P_{x}^{(\alpha_3,\alpha_1)}(u)\!
\left[P_{m}^{(\alpha_2,\alpha_1)}\left(\frac{u+v+2}{u-v}\right) (u-v)^{N}
P_{n}^{(\alpha_4,\alpha_3)}\left(\frac{2-u-v}{v-u}\right)\right]P_{y}^{(\alpha_4,\alpha_2)}(v),\!\!
\label{9j-Integral-2}
\end{gather}
where $h_{n}^{(\alpha,\beta)}$ is given by~\eqref{Norm-Jacobi}.
The integral formula~\eqref{9j-Integral-2} coincides with the one found by Granovskii and
Zhedanov~\cite{Zhedanov-09-1993} using a~related approach.
The formula~\eqref{9j-Integral-2} is one of the most simple expressions for $9j$ symbols.
Given the wealth of results on the asymptotic behavior of Jacobi polynomials, one can expect the
formula~\eqref{9j-Integral-2} to be useful in the examination of the asymptotic behavior of the~$9j$ symbols, an active
f\/ield~\cite{Bonzom-2012,LittleJohn-2011} of interest in particular for the study of spin networks related to quantum
gravity~\cite{Haggard-2010}.

\subsection[The vacuum $9j$ coef\/f\/icients]{The vacuum $\boldsymbol{9j}$ coef\/f\/icients}

The integral expression~\eqref{9j-Integral-2} will now be used to obtain the explicit expression for
the ``va\-cuum''~$9j$ coef\/f\/icients, which correspond to the special case $m=n=0$.
These shall be used in the next section to further characterize the~$9j$ symbols.
Upon using the binomial expansion, the formula~\eqref{9j-Integral-2} gives the following expression for the vacuum~$9j$
coef\/f\/icients:
\begin{gather}
\begin{Bmatrix}
\alpha_1 & \alpha_2 & 0
\\
\alpha_3 & \alpha_4 & 0
\\
x &  y &  N
\end{Bmatrix}
=\left[\frac{c_{0,0;N}}{d_{x,y;N}}
\frac{2^{-N}}{h_{x}^{(\alpha_3,\alpha_1)}h_{y}^{(\alpha_4,\alpha_2)}}\right]
\sum\limits_{k=0}^{N}\binom{N}{k}(-1)^{N-k} \int_{-1}^{1}\int_{-1}^{1}\mathrm{d}u\,
\mathrm{d}v
\nonumber
\\
\qquad{}
\times
(1-u)^{\alpha_3}(1+u)^{\alpha_1}P_{x}^{(\alpha_3,\alpha_1)}(u)u^{k}
(1-v)^{\alpha_4}(1+v)^{\alpha_2} P_{y}^{(\alpha_4,\alpha_2)}(v)v^{N-k},
\label{First}
\end{gather}
where $\binom{N}{k}$ is the binomial coef\/f\/icient.
To evaluate the integrals, one can use the expansion of the power function in series of Jacobi
polynomials~\cite{Andrews_Askey_Roy_1999} which reads
\begin{gather*}
x^{k}=\sum\limits_{j=0}^{k}\left\{\frac{2^{j}k!}{(k-j)!}\,
\frac{\Gamma(j+\alpha+\beta+1)}{\Gamma(2j+\alpha+\beta+1)}\,
\pFq{2}{1}{j-k,j+\alpha+1}{2j+\alpha+\beta+2}{2}\right\}
P_{j}^{(\alpha,\beta)}(x),
\end{gather*}
where ${}_pF_{q}$ stands for the generalized hypergeometric function~\cite{Andrews_Askey_Roy_1999}.
Upon inserting the above expansion in~\eqref{First} and using the orthogonality relation~\eqref{Ortho-Jacobi} for the
Jacobi polynomials, one f\/inds
\begin{gather*}
\begin{Bmatrix}
\alpha_1 & \alpha_2 & 0
\\
\alpha_3 & \alpha_4 & 0
\\
x &  y &  N
\end{Bmatrix}
=\left[\frac{\eta_0^{(\alpha_2,\alpha_1)}\eta_0^{(\alpha_4,\alpha_3)}\eta_{N}^{(\alpha_3+\alpha_4+1,\alpha_1+\alpha_2+1)}}
{\eta_{x}^{(\alpha_3,\alpha_1)}\eta_{y}^{(\alpha_4,\alpha_2)}\eta_{N-x-y}^{(2y+\alpha_2+\alpha_4+1,2x+\alpha_1+\alpha_3+1)}}\right]
\\
\quad{}
\times
\left[\frac{(-1)^{N+x+y}}{2^{N-x-y}}\frac{(N+|\alpha|+3)_{N}}{(N+x+y+|\alpha|+3)_{N-x-y}}\right]\!
\left[\frac{\Gamma(x+\alpha_1\!+\alpha_3\!+1)\Gamma(y+\alpha_2\!+\alpha_4\!+1)}
{\Gamma(2x+\alpha_1\!+\alpha_3\!+1)\Gamma(2y+\alpha_2\!+\alpha_4\!+1)}\right] \\
\quad{}
\times \sum\limits_{k=0}^{N-x-y}\binom{N-x-y}{k}(-1)^{k}
\\
\quad{}
\times \pFq{2}{1}{-k,x+\alpha_3+1}{2x+\alpha_1+\alpha_3+2}{2}
\pFq{2}{1}{-(N-x-y-k),y+\alpha_4+1}{2y+\alpha_2+\alpha_4+2}{2}.
\end{gather*}
The summation in the above relation can be evaluated by means of the formula
\begin{gather*}
\sum\limits_{\ell=0}^{M}\frac{(-N)_{\ell}}{\ell!}\,\pFq{2}{1}{-\ell,a_1}{b_1}{x}\pFq{2}{1}{\ell-N,a_2}{b_2}{x}=x^{N}
\frac{(a_1)_{N}}{(b_1)_{N}}
\pFq{3}{2}{-N,a_2,1-b_1-N}{b_2,1-a_1-N}{1}.
\end{gather*}
Then using identity $(a)_n=\frac{\Gamma(a+n)}{\Gamma(a)}$, the following expression is obtained:
\begin{gather}
\begin{Bmatrix}
\alpha_1 & \alpha_2 & 0
\\
\alpha_3 & \alpha_4 & 0
\\
x &  y &  N
\end{Bmatrix}
=\binom{N}{x,y}^{1/2}[(\alpha_1+1)_{x}(\alpha_2+1)_{y}(\alpha_3+1)_{x}(\alpha_4+1)_{y}]^{1/2}
\nonumber
\\
\qquad{}
\times
\left[\frac{(N+|\alpha|+3)_{x+y}}{(\alpha_1+\alpha_2+2)_{N}(\alpha_3+\alpha_4+2)_{N}}\right]^{1/2}
\left[\frac{(\alpha_1+\alpha_3+1)_{x}(\alpha_1+\alpha_3+2)_{N+x-y}}{(\alpha_1+\alpha_3+1)_{2x}(\alpha_1+\alpha_3+2)_{2x}}\right]^{1/2}
\nonumber
\\
\qquad{}
\times
\left[\frac{(\alpha_2+\alpha_4+1)_{y}(\alpha_2+\alpha_4+2)_{2y}}{(\alpha_2+\alpha_4+1)_{2y}(\alpha_2+\alpha_4+2)_{N-x+y}}\right]^{1/2}
(y+\alpha_4+1)_{N-x-y}
\nonumber
\\
\qquad{}
\times
\pFq{3}{2}{-(N-x-y),-(N-x+y+\alpha_2+\alpha_4+1),x+\alpha_3+1}{-(N-x+\alpha_4),2x+\alpha_1+\alpha_3+2}{1},
\label{Vacuum-9j}
\end{gather}
where $\binom{N}{x,y}=\frac{N!}{x!y!(N-x-y)!}$ stands for the trinomial coef\/f\/icients.
The analogous formula for the~$9j$ coef\/f\/icients of $\mathfrak{su}(2)$ has been given by Hoare and
Rahman~\cite{Hoare-2008}.
The duality formula~\eqref{Duality} can be used to obtain a~similar expression for the case where $x=y=0$.

\section{Raising, lowering operators and contiguity relations}\label{Section4}

In this section, raising and lowering operators are introduced and are called upon to obtain the relations between
contiguous $9j$ symbols by direct computation.
These relations are used to show that the $9j$ symbols can be expressed as the product of the vacuum $9j$ coef\/f\/icients
and a~rational function of the variables $x$, $y$.

\subsection{Raising, lowering operators and factorization}

Let $A_{\pm}^{(\alpha_1,\alpha_2)}$ be def\/ined as
\begin{gather}
\label{Raise-Lower-A}
A_{\pm}^{(\alpha_1,\alpha_2)} =\frac{1}{2} \left[\pm \partial_{\phi_1}- \tg\phi_1(\alpha_1+1/2)+\frac{1}{\tg\phi_1}(\alpha_2+ 1/2)\right],
\end{gather}
and let $B_{\pm}^{(\alpha_3,\alpha_4)}$ have the expression
\begin{gather}
\label{Raise-Lower-B}
B_{\pm}^{(\alpha_3,\alpha_4)} =\frac{1}{2} \left[\pm \partial_{\phi_2}- \tg\phi_2(\alpha_3+1/2)+\frac{1}{\tg\phi_2}(\alpha_4+ 1/2)\right],
\end{gather}
where the coordinates~\eqref{Coords-1} have been used.
It is directly checked that with respect to the scalar product in~\eqref{Ortho-1},
one has $(A_{\pm}^{(\alpha_1,\alpha_2)})^{\dagger}=A_{\mp}^{(\alpha_1,\alpha_2)}$ and
$(B_{\pm}^{(\alpha_3,\alpha_4)})^{\dagger}=B_{\mp}^{(\alpha_3,\alpha_4)}$, where $x^{\dagger}$ stands for the adjoint
of~$x$.
With the help of the relations~\eqref{Raise-Jacobi} and~\eqref{Lower-Jacobi}, it is easily verif\/ied that one has on the
one hand
\begin{gather}
A_{+}^{(\alpha_1,\alpha_2)}\Psi_{m,n;N}^{(\alpha_1+1,\alpha_2+1,\alpha_3,\alpha_4)} (\theta,\phi_1,\phi_2)
\nonumber
\\
\qquad{}
=\sqrt{(m+1)(m+\alpha_1+\alpha_2+2)}\,\Psi_{m+1,n;N+1}^{(\alpha_1,\alpha_2,\alpha_3,\alpha_4)}(\theta,\phi_1,\phi_2),
\nonumber
\\
A_{-}^{(\alpha_1,\alpha_2)}\Psi_{m,n;N}^{(\alpha_1,\alpha_2,\alpha_3,\alpha_4)} (\theta,\phi_1,\phi_2)
\nonumber
\\
\qquad{}
=\sqrt{m(m+\alpha_1+\alpha_2+1)}\,\Psi_{m-1,n;N-1}^{(\alpha_1+1,\alpha_2+1,\alpha_3,\alpha_4)}(\theta,\phi_1,\phi_2),
\label{Action-1}
\end{gather}
and on the other hand
\begin{gather*}
B_{+}^{(\alpha_3,\alpha_4)}\Psi_{m,n;N}^{(\alpha_1,\alpha_2,\alpha_3+1,\alpha_4+1)} (\theta,\phi_1,\phi_2)
\\
\qquad{}
=\sqrt{(n+1)(n+\alpha_3+\alpha_4+2)}\,\Psi_{m,n+1;N+1}^{(\alpha_1,\alpha_2,\alpha_3,\alpha_4)}(\theta,\phi_1,\phi_2),
\\
B_{-}^{(\alpha_3,\alpha_4)}\Psi_{m,n;N}^{(\alpha_1,\alpha_2,\alpha_3,\alpha_4)} (\theta,\phi_1,\phi_2)
\\
\qquad{}
=\sqrt{n(n+\alpha_3+\alpha_4+1)}\,\Psi_{m,n-1;N-1}^{(\alpha_1,\alpha_2,\alpha_3+1,\alpha_4+1)}(\theta,\phi_1,\phi_2),
\end{gather*}
where $\Psi_{m,n;N}^{(\alpha_1,\alpha_2,\alpha_3,\alpha_4)}$ is given by~\eqref{Wave-Scheme-1}.
The operators~\eqref{Raise-Lower-A} and~\eqref{Raise-Lower-B} provide a~factorization of the intermediate Casimir
operators $Q^{(12)}$ and $Q^{(34)}$, respectively.
Indeed, it is directly checked that
\begin{gather}
A_{+}^{(\alpha_1,\alpha_2)}A_{-}^{(\alpha_1,\alpha_2)} =Q^{(12)}-(\alpha_1/2+\alpha_2/2)(\alpha_1/2+\alpha_2/2+1),
\nonumber
\\
B_{+}^{(\alpha_3,\alpha_4)}B_{-}^{(\alpha_3,\alpha_4)} =Q^{(34)}-(\alpha_3/2+\alpha_4/2)(\alpha_3/2+\alpha_4/2+1).
\label{Factorization}
\end{gather}

\subsection{Contiguity relations}

The raising/lowering operators~\eqref{Raise-Lower-A} and~\eqref{Raise-Lower-B} can be used to obtain the relations
satisf\/ied by contiguous $9j$ symbols.
To facilitate the computations, let us make explicit the dependence of the canonical basis vectors $\kket{m,n}{N}$,
$\kket{x,y}{N}$ on the parameters $\alpha_i$ by writing
\begin{gather*}
\kket{m,n}{N}\equiv \kket{\alpha_1,\alpha_2,\alpha_3,\alpha_4;m,n}{N},
\qquad
\kket{x,y}{N}\equiv \kket{\alpha_1,\alpha_2,\alpha_3,\alpha_4;x,y}{N}.
\end{gather*}
With this notation the $9j$ symbols are written as
\begin{gather*}
\begin{Bmatrix}
\alpha_1 & \alpha_2 & m
\\
\alpha_3 & \alpha_4 & n
\\
x & y & N
\end{Bmatrix}
={{}_{N}\langle \alpha_1,\alpha_2,\alpha_3,\alpha_4;x,y|\alpha_1,\alpha_2,\alpha_3,\alpha_4;m,n\rangle_{N}}.
\end{gather*}
To obtain the f\/irst contiguity relation for $9j$ symbols, one considers the matrix element
\begin{gather*}
\BBraket{N}{\alpha_1,\alpha_2,\alpha_3,\alpha_4;x,y}{A_{+}^{(\alpha_1,\alpha_2)}}{\alpha_1+1,\alpha_2+1,\alpha_3,\alpha_4;m,n}{N-1}.
\end{gather*}
By acting with $A_{+}^{(\alpha_1,\alpha_2)}$ on $\kket{\alpha_1+1,\alpha_2+1,\alpha_3,\alpha_4;m,n}{N-1}$
using~\eqref{Action-1}, one f\/inds
\begin{gather}
\label{Rel-1}
\sqrt{(m+1)(m+\alpha_1+\alpha_2+2)}
\begin{Bmatrix}
\alpha_1 &  \alpha_2 & m+1
\\
\alpha_3 & \alpha_4 & n
\\
x & y & N
\end{Bmatrix}
\nonumber
\\
\qquad{}
=\BBraket{N}{\alpha_1,\alpha_2,\alpha_3,\alpha_4;x,y}{A_{+}^{(\alpha_1,\alpha_2)}}{\alpha_1+1,\alpha_2+1,\alpha_3,\alpha_4;m,n}{N-1}.
\end{gather}
To obtain the desired relation, one must determine
$\bbra{\alpha_1,\alpha_2,\alpha_3,\alpha_4;x,y}{N}A_{+}^{(\alpha_1,\alpha_2)}$ or equivalently
\begin{gather*}
\big(A_{+}^{(\alpha_1,\alpha_2)}\big)^{\dagger}\kket{\alpha_1,\alpha_2,\alpha_3,\alpha_4;x,y}{N}
=A_{-}^{(\alpha_1,\alpha_2)}\kket{\alpha_1,\alpha_2,\alpha_3,\alpha_4;x,y}{N},
\end{gather*}
where the reality of the basis functions $\Xi_{x,y;N}$ has been used.
This can be done directly by writing $A_{-}^{(\alpha_1,\alpha_2)}$ in the coordinates
$\{\vartheta,\varphi_1,\varphi_2\}$ def\/ined in~\eqref{Coords-2}, acting with this operator on the wavefunctions
$\Xi_{x,y;N}^{(\alpha_1,\alpha_2,\alpha_3,\alpha_4)}(\vartheta,\varphi_1,\varphi_2)$ and using the properties of the
Jacobi polynomials.
Since this step represents no fundamental dif\/f\/iculties, the details of the computation are relegated to Appendix~\ref{appendixB}. One
f\/inds that
\begin{gather}
A_{-}^{(\alpha_1,\alpha_2)}
\Xi_{x,y;N}^{(\alpha_1,\alpha_2,\alpha_3,\alpha_4)}
\nonumber
\\
{}
=\sqrt{\frac{(x+\alpha_1+1)(x+\alpha_{13}+1)(y+\alpha_2+1)(y+\alpha_{24}+1)(N-x-y)(N+x+y+|\alpha|+3)}
{(2x+\alpha_{13}+1)(2x+\alpha_{13}+2)(2y+\alpha_{24}+1)(2y+\alpha_{24}+2)}\!}\!
\nonumber
\\
{}\times \Xi_{x,y;N-1}^{(\alpha_1+1,\alpha_2+1,\alpha_3,\alpha_4)}
\nonumber
\\
{}
+\sqrt{\frac{x(x+\alpha_3)(y+\alpha_2+1)(y+\alpha_{24}+1)(N-x+y+\alpha_{24}+2)(N+x-y+\alpha_{13}+1)}
{(2x+\alpha_{13})(2x+\alpha_{13}+1)(2y+\alpha_{24}+1)(2y+\alpha_{24}+2)}}
\nonumber
\\
{}\times\Xi_{x-1,y;N-1}^{(\alpha_1+1,\alpha_2+1,\alpha_3,\alpha_4)}
\nonumber
\\
{}-\sqrt{\frac{(x+\alpha_1+1)(x+\alpha_{13}+1)y(y+\alpha_4)(N+x-y+\alpha_{13}+2)(N-x+y+\alpha_{24}+1)}
{(2x+\alpha_{13}+1)(2x+\alpha_{13}+2)(2y+\alpha_{24})(2y+\alpha_{24}+1)}}
\nonumber
\\
{}\times \Xi_{x,y-1;N-1}^{(\alpha_1+1,\alpha_2+1,\alpha_3,\alpha_4)}
\nonumber
\\
{}
-\sqrt{\frac{x(x+\alpha_3)y(y+\alpha_4)(N-x-y+1)(N+x+y+|\alpha|+2)}
{(2x+\alpha_{13})(2x+\alpha_{13}+1)(2y+\alpha_{24})(2y+\alpha_{24}+1)}}
\Xi_{x-1,y-1;N-1}^{(\alpha_1+1,\alpha_2+1,\alpha_3,\alpha_4)},
\label{Rel-2}
\end{gather}
where the shorthand notation $\alpha_{ij}=\alpha_i+\alpha_j$ was used.
Combining~\eqref{Rel-1} with~\eqref{Rel-2}, one f\/inds the contiguity relation
\begin{gather}
\sqrt{(m+1)(m+\alpha_{12}+2)}
\begin{Bmatrix}
\alpha_1 &  \alpha_2 & m+1
\\
\alpha_3 & \alpha_4 & n
\\
x & y & N
\end{Bmatrix}
\nonumber
\\
{}=\sqrt{\frac{(x+\alpha_1+1)(x+\alpha_{13}+1)(y+\alpha_2+1)(y+\alpha_{24}+1)(N-x-y)(N+x+y+|\alpha|+3)}
{(2x+\alpha_{13}+1)(2x+\alpha_{13}+2)(2y+\alpha_{24}+1)(2y+\alpha_{24}+2)}\!}\!
\nonumber
\\
{}\times
\begin{Bmatrix}
\alpha_1+1 &  \alpha_2+1 & m
\\
\alpha_3 & \alpha_4 & n
\\
x & y & N-1
\end{Bmatrix}
\nonumber
\\
{}
+\sqrt{\frac{x(x+\alpha_3)(y+\alpha_2+1)(y+\alpha_{24}+1)(N-x+y+\alpha_{24}+2)(N+x-y+\alpha_{13}+1)}
{(2x+\alpha_{13})(2x+\alpha_{13}+1)(2y+\alpha_{24}+1)(2y+\alpha_{24}+2)}}
\nonumber
\\
{}
\times
\begin{Bmatrix}
\alpha_1+1 &  \alpha_2+1 & m
\\
\alpha_3 & \alpha_4 & n
\\
x-1 & y & N-1
\end{Bmatrix}
\nonumber
\\
{}
-\sqrt{\frac{(x+\alpha_1+1)(x+\alpha_{13}+1)y(y+\alpha_4)(N+x-y+\alpha_{13}+2)(N-x+y+\alpha_{24}+1)}
{(2x+\alpha_{13}+1)(2x+\alpha_{13}+2)(2y+\alpha_{24})(2y+\alpha_{24}+1)}}
\nonumber
\\
{}
\times
\begin{Bmatrix}
\alpha_1+1 &  \alpha_2+1 & m
\\
\alpha_3 & \alpha_4 & n
\\
x & y-1 & N-1
\end{Bmatrix}
\nonumber
\\
{}
-\sqrt{\frac{x(x+\alpha_3)y(y+\alpha_4)(N-x-y+1)(N+x+y+|\alpha|+2)}
{(2x+\alpha_{13})(2x+\alpha_{13}+1)(2y+\alpha_{24})(2y+\alpha_{24}+1)}}
\nonumber
\\
{}
\times
\begin{Bmatrix}
\alpha_1+1 &  \alpha_2+1 & m
\\
\alpha_3 & \alpha_4 & n
\\
x-1 & y-1 & N-1
\end{Bmatrix}.
\label{Contiguous-1}
\end{gather}
To obtain the second contiguity relation, we could consider the matrix element
\begin{gather*}
\BBraket{N}{\alpha_1,\alpha_2,\alpha_3,\alpha_4;x,y}{B_{+}^{(\alpha_3,\alpha_4)}}{\alpha_1,\alpha_2,\alpha_3+1,\alpha_4+1;m,n}{N-1},
\end{gather*}
and proceed similarly by direct computation.
However, it is easier to use the symmetry relation~\eqref{Duality-2} to permute the f\/irst two rows of the
relation~\eqref{Contiguous-1} and then take $\alpha_1\leftrightarrow \alpha_3$, $\alpha_2\leftrightarrow \alpha_4$,
$m\leftrightarrow n$.
This directly leads to the second contiguity relation
\begin{gather}
\sqrt{(n+1)(n+\alpha_{34}+2)}
\begin{Bmatrix}
\alpha_1 &  \alpha_2 & m
\\
\alpha_3 & \alpha_4 & n+1
\\
x & y & N
\end{Bmatrix}
\nonumber
\\
{}
=\sqrt{\frac{(x+\alpha_3+1)(x+\alpha_{13}+1)(y+\alpha_4+1)(y+\alpha_{24}+1)(N-x-y)(N+x+y+|\alpha|+3)}
{(2x+\alpha_{13}+1)(2x+\alpha_{13}+2)(2y+\alpha_{24}+1)(2y+\alpha_{24}+2)}\!}\!
\nonumber
\\
{}
\times
\begin{Bmatrix}
\alpha_1 &  \alpha_2 & m
\\
\alpha_3+1 & \alpha_4+1 & n
\\
x & y & N-1
\end{Bmatrix}
\nonumber
\\
{}
-\sqrt{\frac{x(x+\alpha_1)(y+\alpha_4+1)(y+\alpha_{24}+1)(N-x+y+\alpha_{24}+2)(N+x-y+\alpha_{13}+1)}
{(2x+\alpha_{13})(2x+\alpha_{13}+1)(2y+\alpha_{24}+1)(2y+\alpha_{24}+2)}}
\nonumber
\\
{}
\times
\begin{Bmatrix}
\alpha_1 &  \alpha_2 & m
\\
\alpha_3+1 & \alpha_4+1 & n
\\
x-1 & y & N-1
\end{Bmatrix}
\nonumber
\\
{}
+\sqrt{\frac{(x+\alpha_3+1)(x+\alpha_{13}+1)y(y+\alpha_2)(N+x-y+\alpha_{13}+2)(N-x+y+\alpha_{24}+1)}
{(2x+\alpha_{13}+1)(2x+\alpha_{13}+2)(2y+\alpha_{24})(2y+\alpha_{24}+1)}}
\nonumber
\\
{}
\times
\begin{Bmatrix}
\alpha_1 &  \alpha_2 & m
\\
\alpha_3+1 & \alpha_4+1 & n
\\
x & y-1 & N-1
\end{Bmatrix}
\nonumber
\\
{}
-\sqrt{\frac{x(x+\alpha_1)y(y+\alpha_2)(N-x-y+1)(N+x+y+|\alpha|+2)}
{(2x+\alpha_{13})(2x+\alpha_{13}+1)(2y+\alpha_{24})(2y+\alpha_{24}+1)}}
\nonumber
\\
{}
\times
\begin{Bmatrix}
\alpha_1 &  \alpha_2 & m
\\
\alpha_3+1 & \alpha_4+1 & n
\\
x-1 & y-1 & N-1
\end{Bmatrix}.
\label{Contiguous-2}
\end{gather}
A~third contiguity relation can be found by considering the matrix element
\begin{gather*}
\BBraket{N}{\alpha_1,\alpha_2,\alpha_3,\alpha_4;x,y}{A_{-}^{(\alpha_1-1,\alpha_2-1)}}{m,n;\alpha_1-1,\alpha_2-1,\alpha_3,\alpha_4}{N+1}.
\end{gather*}
Upon using the action~\eqref{Action-1}, one has
\begin{gather*}
\sqrt{m(m+\alpha_1+\alpha_2-1)}
\begin{Bmatrix}
\alpha_1 & \alpha_2 & m-1
\\
\alpha_3 & \alpha_4 & n
\\
x & y & N
\end{Bmatrix}
\\
\qquad
=\BBraket{N}{\alpha_1,\alpha_2,\alpha_3,\alpha_4;x,y}{A_{-}^{(\alpha_1-1,\alpha_2-1)}}{m,n;\alpha_1-1,\alpha_2-1,\alpha_3,\alpha_4}{N}.
\end{gather*}
To obtain the relation, one needs to compute
$\bbra{\alpha_1,\alpha_2,\alpha_3,\alpha_4;x,y}{N}A_{-}^{(\alpha_1-1,\alpha_2-1)}$ or equivalently
\begin{gather}
\label{Dompe}
\big(A_{-}^{(\alpha_1-1,\alpha_2-1)}\big)^{\dagger}\kket{\alpha_1,\alpha_2,\alpha_3,\alpha_4;x,y}{N}
=A_{+}^{(\alpha_1-1,\alpha_2-1)}\kket{\alpha_1,\alpha_2,\alpha_3,\alpha_4;x,y}{N}.
\end{gather}
Following the calculations of Appendix~\ref{appendixC}, one arrives at
\begin{gather}
A_{+}^{(\alpha_1-1,\alpha_2-1)}\Xi_{x,y;N}^{(\alpha_1,\alpha_2,\alpha_3,\alpha_4)}
\nonumber
\\
{}
=\sqrt{\frac{(x+\alpha_1)(x+\alpha_{13})(y+\alpha_2)(y+\alpha_{24})(N-x-y+1)(N+x+y+|\alpha|+2)}
{(2x+\alpha_{13})(2x+\alpha_{13}+1)(2y+\alpha_{24})(2y+\alpha_{24}+1)}}
\nonumber
\\
{}
\times \Xi_{x,y;N+1}^{(\alpha_1-1,\alpha_2-1,\alpha_3,\alpha_4)}
\nonumber
\\
{}
+\sqrt{\frac{(x+1)(x+\alpha_3+1)(y+\alpha_2)(y+\alpha_{24})(N-x+y+\alpha_{24}+1)(N+x-y+\alpha_{13}+2)}
{(2x+\alpha_{13}+1)(2x+\alpha_{13}+2)(2y+\alpha_{24})(2y+\alpha_{24}+1)}}
\nonumber
\\
{}
\times \Xi_{x+1,y;N+1}^{(\alpha_1-1,\alpha_2-1,\alpha_3,\alpha_4)}
\nonumber
\\
{}
-\sqrt{\frac{(x+\alpha_1)(x+\alpha_{13})(y+1)(y+\alpha_4+1)(N+x-y+\alpha_{13}+1)(N-x+y+\alpha_{24}+2)}
{(2x+\alpha_{13})(2x+\alpha_{13}+1)(2y+\alpha_{24}+1)(2y+\alpha_{24}+2)}}
\nonumber
\\
{}
\times \Xi_{x,y+1;N+1}^{(\alpha_1-1,\alpha_2-1,\alpha_3,\alpha_4)}
\nonumber
\\
{}
-\sqrt{\frac{(x+1)(x+\alpha_3+1)(y+1)(y+\alpha_4+1)(N-x-y)(N+x+y+|\alpha|+3)}
{(2x+\alpha_{13}+1)(2x+\alpha_{13}+2)(2y+\alpha_{24}+1)(2y+\alpha_{24}+2)}}
\nonumber
\\
{}
\times \Xi_{x+1,y+1;N+1}^{(\alpha_1-1,\alpha_2-1,\alpha_3,\alpha_4)}.
\label{Action-Ap}
\end{gather}
Combining the above relation with~\eqref{Dompe}, there comes
\begin{gather}
\sqrt{m(m+\alpha_{12}-1)}
\begin{Bmatrix}
\alpha_1 & \alpha_2 & m-1
\\
\alpha_3 & \alpha_4 & n
\\
x & y & N
\end{Bmatrix}
\nonumber
\\
{}
=\sqrt{\frac{(x+\alpha_1)(x+\alpha_{13})(y+\alpha_2)(y+\alpha_{24})(N-x-y+1)(N+x+y+|\alpha|+2)}
{(2x+\alpha_{13})(2x+\alpha_{13}+1)(2y+\alpha_{24})(2y+\alpha_{24}+1)}}
\nonumber
\\
{}
\times
\begin{Bmatrix}
\alpha_1-1 & \alpha_2-1 & m
\\
\alpha_3 & \alpha_4 & n
\\
x & y & N+1
\end{Bmatrix}
\nonumber
\\
{}
+\sqrt{\frac{(x+1)(x+\alpha_3+1)(y+\alpha_2)(y+\alpha_{24})(N-x+y+\alpha_{24}+1)(N+x-y+\alpha_{13}+2)}
{(2x+\alpha_{13}+1)(2x+\alpha_{13}+2)(2y+\alpha_{24})(2y+\alpha_{24}+1)}}
\nonumber
\\
{}
\times
\begin{Bmatrix}
\alpha_1-1 & \alpha_2-1 & m
\\
\alpha_3 & \alpha_4 & n
\\
x+1 & y & N+1
\end{Bmatrix}
\nonumber
\\
{}
-\sqrt{\frac{(x+\alpha_1)(x+\alpha_{13})(y+1)(y+\alpha_4+1)(N+x-y+\alpha_{13}+1)(N-x+y+\alpha_{24}+2)}
{(2x+\alpha_{13})(2x+\alpha_{13}+1)(2y+\alpha_{24}+1)(2y+\alpha_{24}+2)}}
\nonumber
\\
{}
\times
\begin{Bmatrix}
\alpha_1-1 & \alpha_2-1 & m
\\
\alpha_3 & \alpha_4 & n
\\
x & y+1 & N+1
\end{Bmatrix}
\nonumber
\\
{}
-\sqrt{\frac{(x+1)(x+\alpha_3+1)(y+1)(y+\alpha_4+1)(N-x-y)(N+x+y+|\alpha|+3)}
{(2x+\alpha_{13}+1)(2x+\alpha_{13}+2)(2y+\alpha_{24}+1)(2y+\alpha_{24}+2)}}
\nonumber
\\
{}
\times
\begin{Bmatrix}
\alpha_1-1 & \alpha_2-1 & m
\\
\alpha_3 & \alpha_4 & n
\\
x+1 & y+1 & N+1
\end{Bmatrix}.
\label{Contiguous-3}
\end{gather}
Upon applying the symmetry relation~\eqref{Duality-2} on~\eqref{Contiguous-3} and then performing the substitutions
$\alpha_1\leftrightarrow \alpha_3$, $\alpha_2\leftrightarrow \alpha_4$ and $m\leftrightarrow n$, one f\/inds a~fourth
contiguity relation
\begin{gather}
\sqrt{n(n+\alpha_{34}-1)}
\begin{Bmatrix}
\alpha_1 & \alpha_2 & m
\\
\alpha_3 & \alpha_4 & n-1
\\
x & y & N
\end{Bmatrix}
\nonumber
\\
{}
=\sqrt{\frac{(x+\alpha_3)(x+\alpha_{13})(y+\alpha_4)(y+\alpha_{24})(N-x-y+1)(N+x+y+|\alpha|+2)}
{(2x+\alpha_{13})(2x+\alpha_{13}+1)(2y+\alpha_{24})(2y+\alpha_{24}+1)}}
\nonumber
\\
{}
\times
\begin{Bmatrix}
\alpha_1 & \alpha_2 & m
\\
\alpha_3-1 & \alpha_4-1 & n
\\
x & y & N+1
\end{Bmatrix}
\nonumber
\\
{}
-\sqrt{\frac{(x+1)(x+\alpha_1+1)(y+\alpha_4)(y+\alpha_{24})(N-x+y+\alpha_{24}+1)(N+x-y+\alpha_{13}+2)}
{(2x+\alpha_{13}+1)(2x+\alpha_{13}+2)(2y+\alpha_{24})(2y+\alpha_{24}+1)}}
\nonumber
\\
{}
\times
\begin{Bmatrix}
\alpha_1 & \alpha_2 & m
\\
\alpha_3-1 & \alpha_4-1 & n
\\
x+1 & y & N+1
\end{Bmatrix}
\nonumber
\\
{}
+\sqrt{\frac{(x+\alpha_3)(x+\alpha_{13})(y+1)(y+\alpha_2+1)(N+x-y+\alpha_{13}+1)(N-x+y+\alpha_{24}+2)}
{(2x+\alpha_{13})(2x+\alpha_{13}+1)(2y+\alpha_{24}+1)(2y+\alpha_{24}+2)}}
\nonumber
\\
{}
\times
\begin{Bmatrix}
\alpha_1 & \alpha_2 & m
\\
\alpha_3-1 & \alpha_4-1 & n
\\
x & y+1 & N+1
\end{Bmatrix}
\nonumber
\\
{}
-\sqrt{\frac{(x+1)(x+\alpha_1+1)(y+1)(y+\alpha_2+1)(N-x-y)(N+x+y+|\alpha|+3)}{(2x+\alpha_{13}+1)(2x+\alpha_{13}+2)(2y+\alpha_{24}+1)(2y+\alpha_{24}+2)}}
\nonumber
\\
{}
\times
\begin{Bmatrix}
\alpha_1 & \alpha_2 & m
\\
\alpha_3-1 & \alpha_4-1 & n
\\
x+1 & y+1 & N+1
\end{Bmatrix}.
\label{Contiguous-4}
\end{gather}
The relations~\eqref{Contiguous-1},~\eqref{Contiguous-2},~\eqref{Contiguous-3} and~\eqref{Contiguous-4} are usually
obtained by writing the $9j$ symbols in terms of Clebsch--Gordan coef\/f\/icients (given in terms of the Hahn polynomials)
and using the properties of the latter.
In our presentation however, these relations emerge from a~direct computation involving Jacobi polynomials.

\subsection[$9j$ symbols and rational functions]{$\boldsymbol{9j}$ symbols and rational functions}

It will now be shown that the $9j$ symbols of $\mathfrak{su}(1,1)$ can be expressed
as the product of the vacuum coef\/f\/icients and a~rational function.
To this end, let us write the $9j$ symbols as
\begin{gather*}
\begin{Bmatrix}
\alpha_1 & \alpha_2 & m
\\
\alpha_3 & \alpha_4 & n
\\
x & y & N
\end{Bmatrix}
=
\begin{Bmatrix}
\alpha_1 & \alpha_2 & 0
\\
\alpha_3 & \alpha_4 & 0
\\
x & y & N
\end{Bmatrix}
R_{m,n;N}^{(\alpha_1,\alpha_2,\alpha_3,\alpha_4)}(x,y),
\end{gather*}
where $R_{0,0;N}(x,y)\equiv 1$, $R_{-1,n;N}(x,y)=R_{m,-1;N}(x,y)=R_{m,n;-1}(x,y)=0$.
Since the vacuum $9j$ coef\/f\/icients are known explicitly, the contiguity
relations~\eqref{Contiguous-1},~\eqref{Contiguous-2} can be used to generate the functions $R_{m,n;N}(x,y)$.
Upon taking
\begin{gather*}
G_{x,y;N}^{(\alpha_1,\alpha_2,\alpha_3,\alpha_4)}
=\pFq{3}{2}{-(N-x-y),-(N-x+y+\alpha_2+\alpha_4+1),x+\alpha_3+1}{-(N-x+\alpha_4),2x+\alpha_1+\alpha_3+2}{1},
\end{gather*}
using the expression~\eqref{Vacuum-9j} for the vacuum coef\/f\/icients, the relations~\eqref{Contiguous-1}
and~\eqref{Contiguous-2} become
\begin{gather}
\sqrt{\frac{(m+1)(m+\alpha_{12}+2)N(N+\alpha_{12}+2)(N+|\alpha|+3)(\alpha_1+1)(\alpha_2+1)}{(\alpha_{12}+2)(\alpha_{12}+3)(N+\alpha_{34}+1)}}
\nonumber
\\
{}
\times
R_{m+1,n;N}^{(\alpha_1,\alpha_2,\alpha_3,\alpha_4)}(x,y)
=\frac{G^{(\alpha_1+1,\alpha_2+1,\alpha_3,\alpha_4)}_{x,y;N-1}}{G_{x,y;N}^{(\alpha_1,\alpha_2,\alpha_3,\alpha_4)}}
R_{m,n;N-1}^{(\alpha_1+1,\alpha_2+1,\alpha_3,\alpha_4)}(x,y)
\nonumber
\\
{}
\times
\left[\frac{(x+\alpha_1+1)(x+\alpha_{13}+1)(y+\alpha_2+1)(y+\alpha_{24}\!+1)(N-x-y)(N+x+y+|\alpha|+3)}
{(2x+\alpha_{13}+1)(2x+\alpha_{13}+2)(2y+\alpha_{24}+1)(N-x+\alpha_4)}\right]\!
\nonumber
\\
{}
+\left[\frac{x(y+\alpha_2+1)(y+\alpha_2+\alpha_4+1)}{(2y+\alpha_{24}+1)}\right]
\frac{G_{x-1,y;N-1}^{(\alpha_1+1,\alpha_2+1,\alpha_3,\alpha_4)}}{G_{x,y;N}^{(\alpha_1,\alpha_2,\alpha_3,\alpha_4)}}
R_{m,n;N-1}^{(\alpha_1+1,\alpha_2+1,\alpha_3,\alpha_4)}(x-1,y)
\nonumber
\\
{}
-\left[\frac{(x+\alpha_1+1)(x+\alpha_{13}+1)(N+x-y+\alpha_{13}+2)y(y+\alpha_4)(N-x+y+\alpha_{24}+1)}
{(N-x+\alpha_4)(2x+\alpha_{13}+1)(2x+\alpha_{13}+2)(2y+\alpha_{24}+1)}\right]
\nonumber
\\
{}
\times
\frac{G_{x,y-1;N-1}^{(\alpha_1+1,\alpha_2+1,\alpha_3,\alpha_4)}}{G_{x,y;N}^{(\alpha_1,\alpha_2,\alpha_3,\alpha_4)}}
R_{m,n;N-1}^{(\alpha_1+1,\alpha_2+1,\alpha_3,\alpha_4)}(x,y-1)
\nonumber
\\
{}
-\left[\frac{x y(y+\alpha_4)}{(2y+\alpha_{24}+1)}\right]
\frac{G_{x-1,y-1;N-1}^{(\alpha_1+1,\alpha_2+1,\alpha_3,\alpha_4)}}{G_{x,y;N}^{(\alpha_1,\alpha_2,\alpha_3,\alpha_4)}}
R_{m,n;N-1}^{(\alpha_1+1,\alpha_2+1,\alpha_3,\alpha_4)}(x-1,y-1),
\label{Raising-1}
\end{gather}
and
\begin{gather}
\sqrt{\frac{(n+1)(n+\alpha_{34}+2)N(N+\alpha_{34}+2)(N+|\alpha|+3)(\alpha_3+1)(\alpha_4+1)}{(\alpha_{34}+2)(\alpha_{34}+3)(N+\alpha_{12}+1)}}
R_{m,n+1;N}^{(\alpha_1,\alpha_2,\alpha_3,\alpha_4)}
\nonumber
\\
{}
=
\frac{G_{x,y;N-1}^{(\alpha_1,\alpha_2,\alpha_3+1,\alpha_4+1)}}{G_{x,y;N}^{(\alpha_1,\alpha_2,\alpha_3,\alpha_4)}}
R_{m,n;N-1}^{(\alpha_1,\alpha_2,\alpha_3+1,\alpha_4+1)}(x,y)
\nonumber
\\
{}
\times
\left[\frac{(x+\alpha_3+1)(x+\alpha_{13}+1)(y+\alpha_{24}+1)(N-x-y)(N+x+y+|\alpha|+3)}{(2x+\alpha_{13}+1)(2x+\alpha_{13}+2)(2y+\alpha_{24}+1)}\right]
\nonumber
\\
{}
-\left[\frac{x(y+\alpha_{24}+1)(N-x+\alpha_4+1)}{(2y+\alpha_{24}+1)}\right]
\frac{G_{x-1,y;N-1}^{(\alpha_1,\alpha_2,\alpha_3+1,\alpha_4+1)}}{G_{x,y;N}^{(\alpha_1,\alpha_2,\alpha_3,\alpha_4)}}
R_{m,n;N-1}^{(\alpha_1,\alpha_2,\alpha_3+1,\alpha_4+1)}(x-1,y)
\nonumber
\\
{}
+\left[\frac{(x+\alpha_3+1)(x+\alpha_{13}+1)(y)(N+x-y+\alpha_{13}+2)(N-x+y+\alpha_{24}+1)}{(2x+\alpha_{13}+1)(2x+\alpha_{13}+2)(2y+\alpha_{24}+1)}\right]
\nonumber
\\
{}
\times
\frac{G_{x,y-1;N-1}^{(\alpha_1,\alpha_2,\alpha_3+1,\alpha_4+1)}}{G_{x,y;N}^{(\alpha_1,\alpha_2,\alpha_3,\alpha_4)}}
R_{m,n;N-1}^{(\alpha_1,\alpha_2,\alpha_3+1,\alpha_4+1)}(x,y-1)
\nonumber
\\
{}
-\left[\frac{xy(N-x+\alpha_4+1)}{(2y+\alpha_{24}+1)}\right]
\frac{G_{x-1,y-1;N-1}^{(\alpha_1,\alpha_2,\alpha_3+1,\alpha_4+1)}}{G_{x,y;N}^{(\alpha_1,\alpha_2,\alpha_3,\alpha_4)}}
R_{m,n;N-1}^{(\alpha_1,\alpha_2,\alpha_3+1,\alpha_4+1)}(x-1,y-1).
\label{Raising-2}
\end{gather}
From~\eqref{Raising-1} and~\eqref{Raising-2}, one can generate the functions $R_{m,n:N}(x,y)$ recursively.
Writing the f\/irst few cases, one sees that the $R_{m,n}(x,y)$ are rational functions of the variables~$x$,~$y$.
This is in contradiction with the assertion of~\cite{Hoare-2008}, where the functions $R_{m,n}(x,y)$ are claimed to
be polynomials in the variables~$x$,~$y$.
In view of the orthogonality relation~\eqref{9j-Ortho}, the rational functions $R_{m,n}(x,y)$ satisfy the orthogonality
relation
\begin{gather*}
\sum\limits_{\substack{x,y
\\
x+y\leqslant N}}t_{x,y;N}R_{m,n;N}(x,y)R_{m',n'}(x,y)=\delta_{mm'}\delta_{nn'},
\end{gather*}
where the weight function is of the form
\begin{gather*}
t_{x,y;N}=
\begin{Bmatrix}
\alpha_1 & \alpha_2 & 0
\\
\alpha_3 & \alpha_4 & 0
\\
x & y & N
\end{Bmatrix}
^2.
\end{gather*}
It is possible to express the $9j$ symbols of $\mathfrak{su}(1,1)$ in terms of polynomials in the two variab\-les~$x$,~$y$ as
was done by Van der Jeugt in~\cite{VDJ-02-2000}.
However the involved family of polynomials $P_{m,n;N}(x,y)$ is of degree $(N-m,N-n)$ the variables $x(x+\alpha_{13}+1)$
and $y(y+\alpha_{24}+1)$ and hence do not include polynomials whose total degree is less then~$N$.

\section{Dif\/ference equations and recurrence relations}\label{Section5}

In this section, it is shown that the factorization property of the intermediate Casimir operators and the contiguity
relations can be used to exhibit dif\/ference equations and recurrence relations for the $9j$ symbols.

A f\/irst dif\/ference equation can be obtained by considering the matrix element
\begin{gather*}
\BBraket{N}{\alpha_1,\alpha_2,\alpha_3,\alpha_4;x,y}{A_{+}^{(\alpha_1,\alpha_2)}A_{-}^{(\alpha_1,\alpha_2)}}{\alpha_1,\alpha_2,\alpha_3,\alpha_4;m,n}{N}.
\end{gather*}
Using~\eqref{Action-1}, one has on the one hand
\begin{gather*}
\BBraket{N}{\alpha_1,\alpha_2,\alpha_3,\alpha_4;x,y}{A_{+}^{(\alpha_1,\alpha_2)}A_{-}^{(\alpha_1,\alpha_2)}}{\alpha_1,\alpha_2,\alpha_3,\alpha_4;m,n}{N}
\\
\qquad{}
= m(m+\alpha_1+\alpha_2+1)
\begin{Bmatrix}
\alpha_1 &  \alpha_2 & m
\\
\alpha_3 & \alpha_4 & n
\\
x & y &  N
\end{Bmatrix}.
\end{gather*}
Using on the other hand~\eqref{Rel-2} and~\eqref{Action-Ap} to compute
$\bbra{\alpha_1,\alpha_2,\alpha_3,\alpha_4;x,y}{N}A_{+}^{(\alpha_1,\alpha_2)}A_{-}^{(\alpha_1,\alpha_2)}$, one arrives
at the dif\/ference equation
\begin{gather}
m(m+\alpha_{12}+1)
\begin{Bmatrix}
\alpha_1 & \alpha_2 & m
\\
\alpha_3 & \alpha_4 & n
\\
x & y & N
\end{Bmatrix}
=E_{x,y}
\begin{Bmatrix}
\alpha_1 & \alpha_2 & m
\\
\alpha_3 & \alpha_4 & n
\\
x-1 & y-1 & N
\end{Bmatrix}
\nonumber
\\
\qquad
\phantom{=}{}
+E_{x+1,y+1}
\begin{Bmatrix}
\alpha_1 & \alpha_2 & m
\\
\alpha_3 & \alpha_4 & n
\\
x+1 & y+1 N
\end{Bmatrix}
+D_{x,y}
\begin{Bmatrix}
\alpha_1 & \alpha_2 & m
\\
\alpha_3 & \alpha_4 & n
\\
x & y-1 & N
\end{Bmatrix}
\nonumber
\\
\qquad
\phantom{=}{}
+D_{x,y+1}
\begin{Bmatrix}
\alpha_1 & \alpha_2 & m
\\
\alpha_3 & \alpha_4 & n
\\
x & y+1 & N
\end{Bmatrix}
+C_{x,y}
\begin{Bmatrix}
\alpha_1 & \alpha_2 & m
\\
\alpha_3 & \alpha_4 & n
\\
x-1 & y & N
\end{Bmatrix}
\nonumber
\\
\qquad
\phantom{=}{}
+C_{x+1,y}
\begin{Bmatrix}
\alpha_1 & \alpha_2 & m
\\
\alpha_3 & \alpha_4 & n
\\
x+1 & y & N
\end{Bmatrix}
+B_{x+1,y}
\begin{Bmatrix}
\alpha_1 & \alpha_2 & m
\\
\alpha_3 & \alpha_4 & n
\\
x+1 & y-1 & N
\end{Bmatrix}
\nonumber
\\
\qquad
\phantom{=}{}
+B_{x,y+1}
\begin{Bmatrix}
\alpha_1 & \alpha_2 & m
\\
\alpha_3 & \alpha_4 & n
\\
x-1 & y+1 & N
\end{Bmatrix}
+A_{x,y}
\begin{Bmatrix}
\alpha_1 & \alpha_2 & m
\\
\alpha_3 & \alpha_4 & n
\\
x & y & N
\end{Bmatrix}.
\label{Difference-1}
\end{gather}
The coef\/f\/icients are given~by
\begin{gather*}
E_{x,y}=-\textstyle{\sqrt{(N+x+y+|\alpha|+1)(N+x+y+|\alpha|+2)}}
\nonumber
\\
{}
\times \sqrt{(N-x-y+1)(N-x-y+2)}
\sqrt{\frac{x(x+\alpha_1)(x+\alpha_3)(x+\alpha_{13})}{(2x+\alpha_{13}-1)(2x+\alpha_{13})^2(2x+\alpha_{13}+1)}}
\nonumber
\\
{}
\times \sqrt{\frac{y(y+\alpha_2)(y+\alpha_4)(y+\alpha_{24})}{(2y+\alpha_{24}-1)(2y+\alpha_{24})^2(2y+\alpha_{24}+1)}},
\\
D_{x,y}=-\textstyle{\sqrt{(N+x-y+\alpha_{13}+2)(N-x+y+\alpha_{24}+1)}}
\nonumber
\\
{}
\times \sqrt{(N-x-y+1)(N+x+y+|\alpha|+2)}
\sqrt{\frac{y(y+\alpha_2)(y+\alpha_{4})(y+\alpha_{24})}{(2y+\alpha_{24}-1)(2y+\alpha_{24})^2(2y+\alpha_{24}+1)}}
\nonumber
\\
{}
\left[\frac{x(x+\alpha_3)}{(2x+\alpha_{13})(2x+\alpha_{13}+1)}+\frac{(x+\alpha_1+1)(x+\alpha_{13}+1)}{(2x+\alpha_{13}+1)(2x+\alpha_{13}+2)}\right],
\\
C_{x,y}=\sqrt{(N+x-y+\alpha_{13}+1)(N-x+y+\alpha_{24}+2)}
\nonumber
\\
{}
\times \sqrt{(N-x-y+1)(N+x+y+|\alpha|+2)}
\sqrt{\frac{x(x+\alpha_1)(x+\alpha_3)(x+\alpha_{13})}{(2x+\alpha_{13}-1)(2x+\alpha_{13})^2(2x+\alpha_{13}+1)}}
\nonumber
\\
{}
\left[\frac{y(y+\alpha_4)}{(2y+\alpha_{24})(2y+\alpha_{24}+1)}+\frac{(y+\alpha_2+1)(y+\alpha_{24}+1)}{(2y+\alpha_{24}+1)(2y+\alpha_{24}+2)}\right],
\\
B_{x,y}=-\sqrt{(N+x-y+\alpha_{13}+1)(N+x-y+\alpha_{13}+2)(N-x+y+\alpha_{24}+1)}
\nonumber
\\
{}
\sqrt{(N-x+y+\alpha_{24}+2)}
\sqrt{\frac{x(x+\alpha_1)(x+\alpha_3)(x+\alpha_{13})}{(2x+\alpha_{13}-1)(2x+\alpha_{13})^2(2x+\alpha_{13}+1)}}
\nonumber
\\
{}
\times \sqrt{\frac{y(y+\alpha_2)(y+\alpha_{4})(y+\alpha_{24})}{(2y+\alpha_{24}-1)(2y+\alpha_{24})^2(2y+\alpha_{24}+1)}},
\\
A_{x,y}=\Big[\frac{(x+\alpha_1+1)(x+\alpha_{13}+1)y(y+\alpha_4)(N+x-y+\alpha_{13}+2)(N-x+y+\alpha_{24}+1)}
{(2x+\alpha_{13}+1)(2x+\alpha_{13}+2)(2y+\alpha_{24})(2y+\alpha_{24}+1)}
\nonumber
\\
{}
+\frac{x(x+\alpha_3)(y+\alpha_2+1)(y+\alpha_{24}+1)(N+x-y+\alpha_{13}+1)(N-x+y+\alpha_{24}+2)}
{(2x+\alpha_{13})(2x+\alpha_{13}+1)(2y+\alpha_{24}+1)(2y+\alpha_{24}+2)}
\nonumber
\\
{}
+\frac{(x+\alpha_1+1)(x+\alpha_{13}+1)(y+\alpha_2+1)(y+\alpha_{24}+1)(N-x-y)(N+x+y+|\alpha|+3)}
{(2x+\alpha_{13}+1)(2x+\alpha_{13}+2)(2y+\alpha_{24}+1)(2y+\alpha_{24}+2)}
\nonumber
\\
{}
+\frac{x(x+\alpha_3)y(y+\alpha_4)(N-x-y+1)(N+x+y+|\alpha|+2)}{(2x+\alpha_{13})(2x+\alpha_{13}+1)(2y+\alpha_{24})(2y+\alpha_{24}+1)}\Big].
\end{gather*}
 A~second dif\/ference equation is found with the help of the symmetry relation~\eqref{Duality-2}.
It reads
\begin{gather}
n(n+\alpha_{34}+1)
\begin{Bmatrix}
\alpha_1 & \alpha_2 & m
\\
\alpha_3 & \alpha_4 & n
\\
x & y & N
\end{Bmatrix}
=\widetilde{E}_{x,y}
\begin{Bmatrix}
\alpha_1 & \alpha_2 & m
\\
\alpha_3 & \alpha_4 & n
\\
x-1 & y-1 & N
\end{Bmatrix}
\nonumber
\\
{}
+\widetilde{E}_{x+1,y+1}
\begin{Bmatrix}
\alpha_1 & \alpha_2 & m
\\
\alpha_3 & \alpha_4 & n
\\
x+1 & y+1 & N
\end{Bmatrix}
-\widetilde{D}_{x,y}
\begin{Bmatrix}
\alpha_1 & \alpha_2 & m
\\
\alpha_3 & \alpha_4 & n
\\
x & y-1 & N
\end{Bmatrix}
-\widetilde{D}_{x,y+1}
\begin{Bmatrix}
\alpha_1 & \alpha_2 & m
\\
\alpha_3 & \alpha_4 & n
\\
x & y+1 & N
\end{Bmatrix}
\nonumber
\\
{}
-\widetilde{C}_{x,y}
\begin{Bmatrix}
\alpha_1 & \alpha_2 & m
\\
\alpha_3 & \alpha_4 & n
\\
x-1 & y & N
\end{Bmatrix}
-\widetilde{C}_{x+1,y}
\begin{Bmatrix}
\alpha_1 & \alpha_2 & m
\\
\alpha_3 & \alpha_4 & n
\\
x+1 & y & N
\end{Bmatrix}
+\widetilde{B}_{x+1,y}
\begin{Bmatrix}
\alpha_1 & \alpha_2 & m
\\
\alpha_3 & \alpha_4 & n
\\
x+1 & y-1 & N
\end{Bmatrix}
\nonumber
\\
\qquad{}
+\widetilde{B}_{x,y+1}
\begin{Bmatrix}
\alpha_1 & \alpha_2 & m
\\
\alpha_3 & \alpha_4 & n
\\
x-1 & y+1 & N
\end{Bmatrix}
+\widetilde{A}_{x,y}
\begin{Bmatrix}
\alpha_1 & \alpha_2 & m
\\
\alpha_3 & \alpha_4 & n
\\
x & y & N
\end{Bmatrix},
\label{Difference-2}
\end{gather}
where the coef\/f\/icients $\widetilde{E}_{x,y}$, $\widetilde{D}_{x,y}$, $\ldots$, etc.
are obtained from $E_{x,y}$, $D_{x,y}$, $\ldots$ by taking $\alpha_1\leftrightarrow \alpha_3$ and
$\alpha_2\leftrightarrow \alpha_4$.
Given the factorization property~\eqref{Factorization}, the r.h.s.\ of equations~\eqref{Difference-1},~\eqref{Difference-2}
give the action of the intermediate Casimir operators $Q^{(12)}$, $Q^{(34)}$ on the basis where $Q^{(13)}$, $Q^{(24)}$
are diagonal.
Using the duality relation~\eqref{Duality}, it possible to write recurrence relations for the $9j$ symbols which give
the action of the intermediate Casimir operators $Q^{(13)}$, $Q^{(24)}$ on the basis where $Q^{(12)}$, $Q^{(34)}$ are
diagonal.
These relations read
\begin{gather}
x(x+\alpha_{13}+1)
\begin{Bmatrix}
\alpha_1 & \alpha_2 & m
\\
\alpha_3 & \alpha_4 & n
\\
x & y & N
\end{Bmatrix}
=\widehat{E}_{m,n}
\begin{Bmatrix}
\alpha_1 & \alpha_2 & m-1
\\
\alpha_3 & \alpha_4 & n-1
\\
x & y & N
\end{Bmatrix}
\nonumber
\\
{}
+\widehat{E}_{m+1,n+1}
\begin{Bmatrix}
\alpha_1 & \alpha_2 & m+1
\\
\alpha_3 & \alpha_4 & n+1
\\
x & y N
\end{Bmatrix}
+\widehat{D}_{m,n}
\begin{Bmatrix}
\alpha_1 & \alpha_2 & m
\\
\alpha_3 & \alpha_4 & n-1
\\
x & y & N
\end{Bmatrix}
+\widehat{D}_{m,n+1}
\begin{Bmatrix}
\alpha_1 & \alpha_2 & m
\\
\alpha_3 & \alpha_4 & n+1
\\
x & y & N
\end{Bmatrix}
\nonumber
\\
{}
+\widehat{C}_{m,n}
\begin{Bmatrix}
\alpha_1 & \alpha_2 & m-1
\\
\alpha_3 & \alpha_4 & n
\\
x & y & N
\end{Bmatrix}
+\widehat{C}_{m+1,n}
\begin{Bmatrix}
\alpha_1 & \alpha_2 & m+1
\\
\alpha_3 & \alpha_4 & n
\\
x & y & N
\end{Bmatrix}
+\widehat{B}_{m+1,n}
\begin{Bmatrix}
\alpha_1 & \alpha_2 & m+1
\\
\alpha_3 & \alpha_4 & n-1
\\
x & y N
\end{Bmatrix}
\nonumber
\\
{}
+\widehat{B}_{m,n+1}
\begin{Bmatrix}
\alpha_1 & \alpha_2 & m-1
\\
\alpha_3 & \alpha_4 & n+1
\\
x & y & N
\end{Bmatrix}
+\widehat{A}_{m,n}
\begin{Bmatrix}
\alpha_1 & \alpha_2 & m
\\
\alpha_3 & \alpha_4 & n
\\
x & y & N
\end{Bmatrix},
\label{Recurrence-1}
\end{gather}
where $\widehat{E}_{m,n}, \widehat{D}_{m,n}, \ldots$ are obtained from $E_{m,n}, D_{m,n}, \ldots$ by taking
$\alpha_2\leftrightarrow \alpha_3$.
The second recurrence relation is
\begin{gather}
y(y+\alpha_{24}+1)
\begin{Bmatrix}
\alpha_1 & \alpha_2 & m
\\
\alpha_3 & \alpha_4 & n
\\
x & y & N
\end{Bmatrix}
=\check{E}_{m,n}
\begin{Bmatrix}
\alpha_1 & \alpha_2 & m-1
\\
\alpha_3 & \alpha_4 & n-1
\\
x & y & N
\end{Bmatrix}
\nonumber
\\
{}
+\check{E}_{m+1,n+1}
\begin{Bmatrix}
\alpha_1 & \alpha_2 & m+1
\\
\alpha_3 & \alpha_4 & n+1
\\
x & y N
\end{Bmatrix}
-\check{D}_{m,n}
\begin{Bmatrix}
\alpha_1 & \alpha_2 & m
\\
\alpha_3 & \alpha_4 & n-1
\\
x & y & N
\end{Bmatrix}
-\check{D}_{m,n+1}
\begin{Bmatrix}
\alpha_1 & \alpha_2 & m
\\
\alpha_3 & \alpha_4 & n+1
\\
x & y & N
\end{Bmatrix}
\nonumber
\\
{}
-\check{C}_{m,n}
\begin{Bmatrix}
\alpha_1 & \alpha_2 & m-1
\\
\alpha_3 & \alpha_4 & n
\\
x & y & N
\end{Bmatrix}
-\check{C}_{m+1,n}
\begin{Bmatrix}
\alpha_1 & \alpha_2 & m+1
\\
\alpha_3 & \alpha_4 & n
\\
x & y & N
\end{Bmatrix}
+\check{B}_{m+1,n}
\begin{Bmatrix}
\alpha_1 & \alpha_2 & m+1
\\
\alpha_3 & \alpha_4 & n-1
\\
x & y N
\end{Bmatrix}
\nonumber
\\
{}
+\check{B}_{m,n+1}
\begin{Bmatrix}
\alpha_1 & \alpha_2 & m-1
\\
\alpha_3 & \alpha_4 & n+1
\\
x & y & N
\end{Bmatrix}
+\check{A}_{m,n}
\begin{Bmatrix}
\alpha_1 & \alpha_2 & m
\\
\alpha_3 & \alpha_4 & n
\\
x & y & N
\end{Bmatrix},
\label{Recurrence-2}
\end{gather}
where $\check{E}_{m,n}$, $\check{D}_{m,n}$, etc.
are obtained from $E_{m,n}$, $D_{m,n}$, etc, by ef\/fecting the permutation $\sigma=(1243)$ on the parameters
$(\alpha_1,\alpha_2,\alpha_3,\alpha_4)$.
Writing once again the $9j$ symbols as
\begin{gather*}
\begin{Bmatrix}
\alpha_1 & \alpha_2 & m
\\
\alpha_3 & \alpha_4 & n
\\
x & y & N
\end{Bmatrix}
=
\begin{Bmatrix}
\alpha_1 & \alpha_2 & 0
\\
\alpha_3 & \alpha_4 & 0
\\
x &  y &  N
\end{Bmatrix}
R_{m,n}(x,y),
\end{gather*}
and def\/ining
\begin{gather*}
\mathbb{R}_0(x,y) =
\begin{pmatrix}
1
\end{pmatrix}
,
\qquad
\mathbb{R}_1(x,y) =
\begin{pmatrix}
R_{1,0}(x,y)
\\
R_{0,1}(x,y)
\end{pmatrix}
,
\qquad
\mathbb{R}_2(x,y) =
\begin{pmatrix}
R_{2,0}(x,y)
\\
R_{1,1}(x,y)
\\
R_{0,2}(x,y)
\end{pmatrix}
,
\qquad
\dots
\end{gather*}
the recurrence relations~\eqref{Recurrence-1} and~\eqref{Recurrence-2} can be written in matrix form as follows
\begin{gather}
x(x+\alpha_{13}+1) \mathbb{R}_{n}(x,y)=q^{(1)}_{n+2}
\mathbb{R}_{n+2}(x,y)+r_{n+1}^{(1)}
\mathbb{R}_{n+1}(x,y)
\nonumber
\\
\hphantom{x(x+\alpha_{13}+1) \mathbb{R}_{n}(x,y)=}{}
+s_{n}^{(1)}
\mathbb{R}_{n}(x,y)+r_{n}^{(1)}
\mathbb{R}_{n-1}(x,y)+q^{(1)}_{n}
\mathbb{R}_{n-2}(x,y),
\label{Dompe-a}
\\
y(y+\alpha_{24}+1) \mathbb{R}_{n}(x,y)=q^{(2)}_{n+2}
\mathbb{R}_{n+2}(x,y)+r_{n+1}^{(2)}
\mathbb{R}_{n+1}(x,y)
\nonumber
\\
\hphantom{y(y+\alpha_{24}+1) \mathbb{R}_{n}(x,y)=}{}
+s_{n}^{(2)}
\mathbb{R}_{n}(x,y)+r_{n}^{(2)}
\mathbb{R}_{n-1}(x,y)+q^{(2)}_{n}
\mathbb{R}_{n-2}(x,y),
\label{Dompe-b}
\end{gather}
where the matrices $q_{n}^{(i)}$, $r^{(i)}_{n}$ and $s^{(i)}_{n}$ are easily found from the coef\/f\/icients
in~\eqref{Recurrence-1} and~\eqref{Recurrence-2}.
It is apparent from~\eqref{Dompe-a} and~\eqref{Dompe-b} that the vector functions $\mathbb{R}_{m}(x,y)$ satisfy a~f\/ive
term recurrence relation.
In view of the multivariate extension of Favard's theorem~\cite{Dunkl-2001}, this conf\/irms that the functions
$\mathbb{R}_{m}(x,y)$ are not orthogonal polynomials.

\section{Conclusion}

In this paper, we have used the connection between the addition of four $\mathfrak{su}(1,1)$ representations of the
positive discrete series and the generic superintegrable model on the 3-sphere to study the $9j$ coef\/f\/icients in the
position representation.
We constructed the canonical basis vectors of the $9j$ problem explicitly and related them to the separation of
variables in cylindrical coordinates.
Moreover, we have obtained by direct computation the contiguity relations, the dif\/ference equations and the recurrence
relations satisf\/ied by the $9j$ symbols.
The properties of the $9j$ coef\/f\/icients as bivariate functions have thus been clarif\/ied.

The present work suggests many avenues for further investigations.
For example Lievens and Van der Jeugt~\cite{Lievens-2002} have constructed explicitly the coupled basis vectors arising
in the tensor product of an arbitrary number of $\mathfrak{su}(1,1)$ representations in the coherent state
representation.
Given this result, it would be of interest to give the realization of these vectors in the position representation~by
examining the generic superintegrable system on the~$n$-sphere.
Another interesting question is that of the orthogonal polynomials in two variables connected with the $9j$ problem.
With the observations of the present work and those made by Van der Jeugt in~\cite{VDJ-1997}, one must conclude that the study
of $9j$ symbols do not naturally lead to families of biva\-riate orthogonal polynomials that would be two-variable
extensions of the Racah polynomials.
However, the results obtained by Kalnins, Miller and Post~\cite{Kalnins-2011-05} and the connection between the generic
model on the three-sphere and the $9j$ problem exhibited here suggest that an algebraic interpretation for the bivariate
extension of the Racah polynomials, as def\/ined by Tratnik~\cite{Tratnik-1991-04}, could be given in the framework of the
addition of four $\mathfrak{su}(1,1)$ algebras by investigating the overlap coef\/f\/icients between bases which are
dif\/ferent from the canonical ones.
We plan to follow up on this.

\appendix

\section{Properties of Jacobi polynomials}\label{appendixA}

The Jacobi polynomials, denoted by $P_{n}^{(\alpha,\beta)}(z)$, are def\/ined as follows~\cite{Koekoek-2010}:
\begin{gather*}
P_{n}^{(\alpha,\beta)}(z)=\frac{(\alpha+1)_{n}}{n!}\pFq{2}{1}{-n,n+\alpha+\beta+1}{\alpha+1}{\frac{1-z}{2}},
\end{gather*}
where ${}_pF_{q}$ stands for the generalized hypergeometric function~\cite{Andrews_Askey_Roy_1999}.
The polynomials satisfy
\begin{gather}
\label{Ortho-Jacobi}
\int_{-1}^{1}(1-z)^{\alpha}(1+z)^{\beta}P_{n}^{(\alpha,\beta)}(z)P_{m}^{(\alpha,\beta)}(z)
\mathrm{d}z=h_{n}^{(\alpha,\beta)}\,\delta_{nm},
\end{gather}
where the normalization coef\/f\/icient is
\begin{gather}
\label{Norm-Jacobi}
h_{n}^{(\alpha,\beta)}
=2^{\alpha+\beta+1}\frac{\Gamma(2n+\alpha+\beta+1)\Gamma(n+\alpha+1)\Gamma(n+\beta+1)}{\Gamma(2n+\alpha+\beta+2)\Gamma(n+\alpha+\beta+1)\Gamma(n+1)}.
\end{gather}
The derivatives of the Jacobi polynomials give~\cite{Miller-1968}
\begin{gather}
\label{Lower-Jacobi}
\partial_{z}P_{n}^{(\alpha,\beta)}(z) =\left[\frac{n+\alpha+\beta+1}{2}\right]
P_{n-1}^{(\alpha+1,\beta+1)}(z),
\\
\label{Raise-Jacobi}
\partial_{z}\left((1-z)^{\alpha}(1+z)^{\beta}P_{n}^{(\alpha,\beta)}(z)\right) =-2(n+1)
(1-z)^{\alpha-1}(1+z)^{\beta-1}
P_{n+1}^{(\alpha-1,\beta-1)}(z).
\end{gather}
One has
\begin{gather}
\label{Raise-Beta-Jacobi}
P_{n}^{(\alpha,\beta)}(z)=\left(\frac{n+\alpha+\beta+1}{2n+\alpha+\beta+1}\right)
P_{n}^{(\alpha,\beta+1)}(z)+\left(\frac{n+\alpha}{2n+\alpha+\beta+1}\right)P_{n-1}^{(\alpha,\beta+1)}(z).
\end{gather}
and
\begin{gather}
\left(\frac{1-z}{2}\right)P_{n-1}^{(\alpha,\beta)}(z)
\nonumber
\\
\qquad
=
\left(\frac{n+\alpha-1}{2n+\alpha+\beta-1}\right)P_{n-1}^{(\alpha_1-1,\beta)}(z)-\left(\frac{n}{2n+\alpha+\beta-1}\right)
P_{n}^{(\alpha-1,\beta)}(z).
\label{Lower-Alpha-Jacobi}
\end{gather}
Since $P_{n}^{(\alpha,\beta)}(-z)=(-1)^{n}P_{n}^{(\beta,\alpha)}(z)$, one has also
\begin{gather}
\left(\frac{1+z}{2}\right)P_{n}^{(\alpha,\beta)}(z)
\nonumber
\\
\qquad{}
=\left(\frac{n+\beta}{2n+\alpha+\beta+1}\right)P_{n}^{(\alpha,\beta-1)}(z)+\left(\frac{n+1}{2n+\alpha+\beta+1}\right)P_{n+1}^{(\alpha,\beta-1)}(z),
\label{Lower-Beta-Jacobi}
\end{gather}
and
\begin{gather}
\label{Raise-Alpha-Jacobi}
P_{n}^{(\alpha,\beta)}(z)
=\left(\frac{n+\alpha+\beta+1}{2n+\alpha+\beta+1}\right)P_{n}^{(\alpha+1,\beta)}(z)
-\left(\frac{n+\beta}{2n+\alpha+\beta+1}\right)P_{n-1}^{(\alpha+1,\beta)}(z).
\end{gather}

\section[Action of $A_{-}^{(\alpha_1,\alpha_2)}$ on $\Xi_{x,y;N}^{(\alpha_1,\alpha_2,\alpha_3,\alpha_4)}$]{Action of $\boldsymbol{A_{-}^{(\alpha_1,\alpha_2)}}$ on $\boldsymbol{\Xi_{x,y;N}^{(\alpha_1,\alpha_2,\alpha_3,\alpha_4)}}$}\label{appendixB}

In Cartesian coordinates, the operator $A_{-}^{(\alpha_1,\alpha_2)}$ reads
\begin{gather*}
A_{-}^{(\alpha_1,\alpha_2)}=-\frac{1}{2}(s_1\partial_{s_2}-s_2\partial_{s_1})+\frac{s_1}{2s_2}(\alpha_2+1/2)-\frac{s_2}{2s_1}(\alpha_1+1/2).
\end{gather*}
The action of $A_{-}^{(\alpha_1,\alpha_2)}$ on the wavefunctions $\Xi_{x,y;N}^{(\alpha_1,\alpha_2,\alpha_3,\alpha_4)}$
can be written as
\begin{gather*}
\mathcal{F}
\eta_{x}^{(\alpha_3,\alpha_1)}\eta_{y}^{(\alpha_4,\alpha_2)}\eta_{N-x-y}^{(2y+\alpha_{24}+1,2x+\alpha_{13}+1)}
\\
\qquad
\times
\left[\mathcal{F}^{-1}A_{-}^{(\alpha_1,\alpha_2)}\mathcal{F}\right]\left[P_{N-x-y}^{(2y+\alpha_{24}+1,2x+\alpha_{13}+1)}(\cos
2\vartheta)P_{x}^{(\alpha_3,\alpha_1)}\left(\cos 2\varphi_1\right)P_{y}^{(\alpha_4,\alpha_2)}\left(\cos
2\varphi_2\right)\right],
\end{gather*}
where
\begin{gather*}
\mathcal{F}=\big(s_1^2+s_3^2\big)^{x}\big(s_2^2+s_4^2\big)^{y}\prod\limits_{i=1}^{4}s_i^{\alpha_i+1/2}.
\end{gather*}
One has
\begin{gather*}
\big[\mathcal{F}^{-1}A_{-}^{(\alpha_1,\alpha_2)}\mathcal{F}\big]
=-\frac{1}{2}(s_1\partial_{s_2}-s_2\partial_{s_1})+x\frac{s_1s_2}{s_1^2+s_3^2}-y\frac{s_1s_2}{s_2^2+s_4^2}.
\end{gather*}
In the cylindrical coordinates~\eqref{Coords-2}, the operator reads
\begin{gather*}
[\mathcal{F}^{-1}A_{-}^{(\alpha_1,\alpha_2)}\mathcal{F}]=x\left[\tg\vartheta
\cos\varphi_1\cos\varphi_2\right]-y\left[\frac{\cos\varphi_1\cos\varphi_2}{\tg\vartheta}\right]
\\
\hphantom{[\mathcal{F}^{-1}A_{-}^{(\alpha_1,\alpha_2)}\mathcal{F}]=}{}
-\frac{1}{2}\left[\cos\varphi_1\cos\varphi_2\,\partial_{\vartheta}+\tg
\vartheta\sin\varphi_1\cos\varphi_2\,\partial_{\varphi_1}-\frac{\cos\varphi_1\sin\varphi_2}{\tg\vartheta}\partial_{\varphi_2}\right].
\end{gather*}
Using the relation~\eqref{Lower-Jacobi}, one f\/inds
\begin{gather*}
A_{-}^{(\alpha_1,\alpha_2)}\Xi_{x,y;N}^{(\alpha_1,\alpha_2,\alpha_3,\alpha_4)}=\upsilon_{x,y,N}^{(\alpha_1,\alpha_2,\alpha_3)}
\Big[(N+x+y+|\alpha|+3)\big(\cos^2\vartheta\big)^{x}\big(\sin^2\vartheta\big)^{y}
\\
\qquad{}
\times P_{N-x-y-1}^{(2y+\alpha_{24}+2,2x+\alpha_{13}+2)}(\cos 2\vartheta)P_{x}^{(\alpha_3,\alpha_1)}(\cos
2\varphi_1)P_{y}^{(\alpha_4,\alpha_2)}(\cos 2\varphi_2)
\\
\qquad{}
+(x+\alpha_{13}+1)\big(\cos^2\vartheta\big)^{x-1}\big(\sin^2\vartheta\big)^{y} \sin^2\varphi_1
\\
\qquad{}
\times P_{N-x-y}^{(2y+\alpha_{24}+1,2x+\alpha_{13}+1)}(\cos 2\vartheta)P_{x-1}^{(\alpha_3+1,\alpha_1+1)}(\cos
2\varphi_1)P_{y}^{(\alpha_4,\alpha_2)}(\cos 2\varphi_2)
\\
\qquad{}
-(y+\alpha_{24}+1)\big(\cos^2\vartheta\big)^{x}\big(\sin^2\vartheta\big)^{y-1}\sin^2\varphi_2
\\
\qquad{}
\times P_{N-x-y}^{(2y+\alpha_{24}+1,2x+\alpha_{13}+1)}(\cos 2\vartheta)P_{x}^{(\alpha_3,\alpha_1)}(\cos
2\varphi_1)P_{y-1}^{(\alpha_4+1,\alpha_2+1)}(\cos 2\varphi_2)
\\
\qquad{}
+\Big[x\big(\cos^2\vartheta\big)^{x-1}\big(\sin^2\vartheta\big)^{y}-y\big(\cos^2\vartheta\big)^{x}\big(\sin^2\vartheta\big)^{y-1}\Big]
\\
\qquad{}
\times P_{N-x-y}^{(2y+\alpha_{24}+1,2x+\alpha_{13}+1)}(\cos 2\vartheta)P_{x}^{(\alpha_3,\alpha_1)}(\cos
2\varphi_1)P_{y}^{(\alpha_4,\alpha_2)}(\cos 2\varphi_2)\Big],
\end{gather*}
where
\begin{gather*}
\upsilon_{x,y,N}^{(\alpha_1,\alpha_2,\alpha_3)}
=\eta_{x}^{(\alpha_3,\alpha_1)}\eta_{y}^{(\alpha_4,\alpha_2)}\eta_{N-x-y}^{(2y+\alpha_{24}+1,2x+\alpha_{13}+1)}
(s_1)^{\alpha_1+3/2}(s_2)^{\alpha_2+3/2}(s_3)^{\alpha_3+1/2}(s_4)^{\alpha_4+1/2}.
\end{gather*}
The identities~\eqref{Raise-Beta-Jacobi} and~\eqref{Lower-Alpha-Jacobi} can then be used to write the result in a~form
involving only terms of the type $P_{k}^{(\alpha_3,\alpha_1+1)}$ and $P_{k'}^{(\alpha_4,\alpha_2+1)}$.
Regrouping the terms, one f\/inds
\begin{gather*}
A_{-}^{(\alpha_1,\alpha_2)}\Xi_{x,y;N}^{(\alpha_1,\alpha_2,\alpha_3,\alpha_4)}=\upsilon_{x,y,N}^{(\alpha_1,\alpha_2,\alpha_3)}
\Bigg[\left\{\frac{(x+\alpha_{13}+1)(y+\alpha_{24}+1)(N+x+y+|\alpha|+3)}{(2x+\alpha_{13}+1)(2y+\alpha_{24}+1)}\right\}
\\
{}
\times \big(\cos^2\vartheta\big)^{x}\big(\sin^2\vartheta\big)^{y} P_{x}^{(\alpha_3,\alpha_1+1)}(\cos
2\varphi_1)P_{y}^{(\alpha_4,\alpha_2+1)}(\cos 2\varphi_2)P_{N-x-y-1}^{(2y+\alpha_{24}+2,2x+\alpha_{13}+2)}(\cos 2\vartheta)
\\
{}
+\left\{\frac{(x+\alpha_3)(y+\alpha_{24}+1)(\cos^2\vartheta)^{x-1}(\sin^2\vartheta)^{y}}{(2x+\alpha_{13}+1)(2y+\alpha_{24}+1)}\right\}
P_{x-1}^{(\alpha_3,\alpha_1+1)}(\cos 2\varphi_1)P_{y}^{(\alpha_4,\alpha_2+1)}(\cos 2\varphi_2)
\\
{}
\times \Big((N+x+y+|\alpha|+3)\cos^2\vartheta P_{N-x-y-1}^{(2y+\alpha_{24}+2,2x+\alpha_{13}+2)}(\cos 2\vartheta)
\\
{}
+(2x+\alpha_{13}+1)P_{N-x-y}^{(2y+\alpha_{24}+1,2x+\alpha_{13}+1)}(\cos 2\vartheta)\Big)
\\
{}
+\left\{\frac{(y+\alpha_4)(x+\alpha_{13}+1)(\cos^2\vartheta)^{x}(\sin^2\vartheta)^{y-1}}{(2x+\alpha_{13}+1)(2y+\alpha_{24}+1)}\right\}
P_{x}^{(\alpha_3,\alpha_1+1)}(\cos 2\varphi_1)P_{y-1}^{(\alpha_4,\alpha_2+1)}(\cos 2\varphi_2)
\\
{}
\times \Big((N+x+y+|\alpha|+3)\sin^2\vartheta P_{N-x-y-1}^{(2y+\alpha_{24}+2,2x+\alpha_{13}+2)}(\cos 2\vartheta)
\\
{}
-(2y+\alpha_{24}+1)P_{n-x-y}^{(2y+\alpha_{24}+1,2x+\alpha_{13}+1)}(\cos 2\vartheta)\Big)
\\
{}
+\left\{\frac{(y+\alpha_4)(x+\alpha_3)(\cos^2\vartheta)^{x-1}(\sin^2\vartheta)^{y-1}}{(2x+\alpha_{13}+1)(2y+\alpha_{24}+1)}\right\}
P_{x-1}^{(\alpha_3,\alpha_1+1)}(\cos 2\varphi_1)P_{y-1}^{(\alpha_4,\alpha_2+1)}(\cos 2\varphi_2)
\\
{}
\times \Big((N+x+y+|\alpha|+3)\cos^2\vartheta\sin^2\vartheta P_{N-x-y-1}^{(2y+\alpha_{24}+2,2x+\alpha_{13}+2)}(\cos
2\vartheta)
\\
{}
+(2x+\alpha_{13}+1)\sin^2\vartheta P_{N-x-y}^{(2y+\alpha_{24}+1,2x+\alpha_{13}+1)}(\cos 2\vartheta)
\\
{}
-(2y+\alpha_{24}+1)\cos^2\vartheta
P_{N-x-y}^{(2y+\alpha_{24}+1,2x+\alpha_{13}+1)}(\cos 2\vartheta)\Big)\bigg].
\end{gather*}
The terms between parentheses in the above expression are easily evaluated and found to be
\begin{gather*}
(N+x+y+|\alpha|+3)
\cos^2\vartheta
P_{N-x-y-1}^{(2y+\alpha_{24}+2,2x+\alpha_{13}+2)}(\cos 2\vartheta)
\\
\qquad
\phantom{=}
+(2x+\alpha_{13}+1)P_{N-x-y}^{(2y+\alpha_{24}+1,2x+\alpha_{13}+1)}(\cos 2\vartheta)
\\
\qquad{}
= (N+x-y+\alpha_{13}+1)P_{N-x-y}^{(2y+\alpha_{24}+2,2x+\alpha_{13})}(\cos 2\vartheta),
\\
(N+x+y+|\alpha|+3)
\sin^2\vartheta
P_{N-x-y-1}^{(2y+\alpha_{24}+2,2x+\alpha_{13}+2)}(\cos 2\vartheta)
\\
\qquad
\phantom{=}{}
-(2y+\alpha_{24}+1)P_{n-x-y}^{(2y+\alpha_{24}+1,2x+\alpha_{13}+1)}(\cos 2\vartheta)
\\
\qquad{}
=-(N-x+y+\alpha_{24}+1)P_{N-x-y}^{(2y+\alpha_{24},2x+\alpha_{13}+2)}(\cos 2\vartheta),
\\
(N+x+y+|\alpha|+3)\cos^2\vartheta\sin^2\vartheta P_{N-x-y-1}^{(2y+\alpha_{24}+2,2x+\alpha_{13}+2)}(\cos 2\vartheta)
\\
\qquad
\phantom{=}{}
+(2x+\alpha_{13}+1)\sin^2\vartheta P_{N-x-y}^{(2y+\alpha_{24}+1,2x+\alpha_{13}+1)}(\cos 2\vartheta)
\\
\qquad
\phantom{=}{}
-(2y+\alpha_{24}+1)\cos^2\vartheta
P_{N-x-y}^{(2y+\alpha_{24}+1,2x+\alpha_{13}+1)}(\cos 2\vartheta)
\\
\qquad{}
=-(N-x-y+1)P_{N-x-y+1}^{(2y+\alpha_{24},2x+\alpha_{13})}(\cos 2\vartheta).
\end{gather*}
Adjusting the normalization factors then yields the result~\eqref{Rel-2}.

\section[Action of $A_{+}^{(\alpha_1-1,\alpha_2-1)}$ on $\Xi_{x,y;N}^{(\alpha_1,\alpha_2,\alpha_3,\alpha_4)}$]{Action of $\boldsymbol{A_{+}^{(\alpha_1-1,\alpha_2-1)}}$ on $\boldsymbol{\Xi_{x,y;N}^{(\alpha_1,\alpha_2,\alpha_3,\alpha_4)}}$}\label{appendixC}

In Cartesian coordinates, the operator $A_{+}^{(\alpha_1-1,\alpha_2-1)}$ reads
\begin{gather*}
A_{+}^{(\alpha_1-1,\alpha_2-1)}=\frac{1}{2}(s_1\partial_{s_2}-s_2\partial_{s_1})+\frac{s_1}{2s_2}(\alpha_2-1/2)-\frac{s_2}{2s_1}(\alpha_1-1/2).
\end{gather*}
The action of $A_{+}^{(\alpha_1-1,\alpha_2-1)}$ on the wavefunctions $\Xi_{x,y;N}$ can be expressed as
\begin{gather*}
\mathcal{G}^{-1}\eta_{x}^{(\alpha_3,\alpha_1)}\eta_{y}^{(\alpha_4,\alpha_2)}\eta_{N-x-y}^{(2y+\alpha_{24}+1,2x+\alpha_{13}+1)}
\\
\qquad
\times
\left[\mathcal{G}A_{+}^{(\alpha_1-1,\alpha_2-1)}\mathcal{G}^{-1}\right]
\left[(\sin^{2}\varphi_1)^{\alpha_3}(\cos^2\varphi_1)^{\alpha_1}P_{x}^{(\alpha_3,\alpha_1)}(\cos
2\varphi_1)\right]
\\
\qquad{}
\times \left[(\sin^{2}\varphi_2)^{\alpha_4}(\cos^2\varphi_2)^{\alpha_2}P_{y}^{(\alpha_4,\alpha_2)}(\cos
2\varphi_2)\right]
\\
\qquad{}
\times
\left[\big(\sin^2\vartheta\big)^{2x+\alpha_{24}+1}\big(\cos^2\vartheta\big)^{2x+\alpha_{13}+1}P_{N-x-y}^{(2y+\alpha_{24}+1,2x+\alpha_{13}+1)}(\cos
2\vartheta)\right],
\end{gather*}
where
\begin{gather*}
\mathcal{G}=\big(s_1^2+s_3^2\big)^{x+1}\big(s_2^2+s_4^2\big)^{y+1}\prod\limits_{i=1}^{4}s_i^{\alpha_i-1/2}.
\end{gather*}
One has
\begin{gather*}
\big[\mathcal{G}A_{+}^{(\alpha_1-1,\alpha_2-1)}\mathcal{G}^{-1}\big]
=\frac{1}{2}(s_2\partial_{s_1}-s_1\partial_{s_2})+(x+1)\frac{s_1s_2}{s_1^2+s_3^2}-(y+1)\frac{s_1s_2}{s_2^2+s_4^2}.
\end{gather*}
In the cylindrical coordinates~\eqref{Coords-2}, this operator reads
\begin{gather*}
\big[\mathcal{G}A_{+}^{(\alpha_1-1,\alpha_2-1)}\mathcal{G}^{-1}\big]=\frac{1}{2}\Big[\cos \varphi_1\cos \varphi_2
\partial_{\vartheta}+\tg\vartheta \sin\varphi_1\cos \varphi_2
\partial_{\varphi_1}
\\
\qquad{}
-\frac{\cos\varphi_1\sin\varphi_2}{\tg\vartheta}\partial_{\varphi_2}\Big]+(x+1)[\tg\vartheta
\cos\varphi_1\cos\varphi_2]-(y+1)
[\mathrm{ctg}\,\vartheta \cos \varphi_1\cos \varphi_2].
\end{gather*}
Using the identity~\eqref{Raise-Jacobi}, one f\/inds
\begin{gather*}
A_{+}^{(\alpha_1-1,\alpha_2-1)}\Xi_{x,y;N}^{(\alpha_1,\alpha_2,\alpha_3,\alpha_4)}=\kappa_{x,y;N}^{(\alpha_1,\alpha_2,\alpha_3)}
\Big[(N-x-y+1)\cos^2\varphi_1\cos^2\varphi_2
\\
\qquad{}
\times \big(\cos^2\vartheta\big)^{x}\big(\sin^2\vartheta\big)^{y} P_{x}^{(\alpha_3,\alpha_1)}(\cos
2\varphi_1)P_{y}^{(\alpha_4,\alpha_2)}(\cos 2\varphi_2)P_{N-x-y+1}^{(2y+\alpha_{24},2x+\alpha_{13})}(\cos 2\vartheta)
\\
\qquad{}
+(x+1)\cos^2\varphi_2\big(\cos^2\vartheta\big)^{x}\big(\sin^2\vartheta\big)^{y+1}
\\
\qquad{}
\times P_{x+1}^{(\alpha_3-1,\alpha_1-1)}(\cos 2\varphi_1)P_{y}^{(\alpha_4,\alpha_2)}(\cos
2\varphi_2)P_{N-x-y}^{(2y+\alpha_{24}+1,2x+\alpha_{13}+1)}(\cos 2\vartheta)
\\
\qquad{}
-(y+1)\cos^2\varphi_1\big(\cos^2\vartheta\big)^{x+1}\big(\sin^2\vartheta\big)^{y}
\\
\qquad{}
\times P_{x}^{(\alpha_3,\alpha_1)}(\cos 2\varphi_1)P_{y+1}^{(\alpha_4-1,\alpha_2-1)}(\cos
2\varphi_2)P_{N-x-y}^{(2y+\alpha_{24}+1,2x+\alpha_{13}+1)}(\cos 2\vartheta)
\\
\qquad{}
+(x+1)\cos^2\varphi_1\cos^2 \varphi_2\big(\cos^2\vartheta\big)^{x}\big(\sin^2\vartheta\big)^{y+1}
\\
\qquad{}
\times P_{x}^{(\alpha_3,\alpha_1)}(\cos 2\varphi_1)P_{y}^{(\alpha_4,\alpha_2)}(\cos
2\varphi_2)P_{N-x-y}^{(2y+\alpha_{24}+1,2x+\alpha_{13}+1)}(\cos 2\vartheta)
\\
\qquad{}
-(y+1)\cos^2\varphi_1\cos^2 \varphi_2\big(\cos^2\vartheta\big)^{x+1}\big(\sin^2\vartheta\big)^{y}
\\
\qquad{}
\times P_{x}^{(\alpha_3,\alpha_1)}(\cos 2\varphi_1)P_{y}^{(\alpha_4,\alpha_2)}(\cos
2\varphi_2)P_{N-x-y}^{(2y+\alpha_{24}+1,2x+\alpha_{13}+1)}(\cos 2\vartheta)\Big],
\end{gather*}
where
\begin{gather*}
\kappa_{x,y;N}^{(\alpha_1,\alpha_2,\alpha_3)}
=\eta_{x}^{(\alpha_3,\alpha_1)}\eta_{y}^{(\alpha_4,\alpha_2)}\eta_{N-x-y}^{(2y+\alpha_{24}+1,2x+\alpha_{13}+1)}
(s_1)^{\alpha_1-1/2}(s_2)^{\alpha_2-1/2}(s_3)^{\alpha_3+1/2}(s_4)^{\alpha_4+1/2}.
\end{gather*}
Then using~\eqref{Lower-Beta-Jacobi} and~\eqref{Raise-Alpha-Jacobi}, one f\/inds
\begin{gather*}
A_{+}^{(\alpha_1-1,\alpha_2-1)}\Xi_{x,y;N}^{(\alpha_1,\alpha_2,\alpha_3,\alpha_4)}=\kappa_{x,y;N}^{(\alpha_1,\alpha_2,\alpha_3)}
\Bigg[\left\{\frac{(x+\alpha_1)(y+\alpha_2)(N-x-y+1)}{(2x+\alpha_{13}+1)(2y+\alpha_{24}+1)}\right\}
\\
\qquad{}
\times \big(\cos^2\vartheta\big)^{x}\big(\sin^2\vartheta\big)^{y}P_{x}^{(\alpha_3,\alpha_1-1)}(\cos
2\varphi_1)P_{y}^{(\alpha_4,\alpha_2-1)}(\cos 2\varphi_2)P_{N-x-y+1}^{(2y+\alpha_{24},2x+\alpha_{13})}(\cos 2\vartheta)
\\
\qquad{}
+\left\{\frac{(x+1)(y+\alpha_2)(\cos^2\vartheta)^{x+1}(\sin^2\vartheta)^{y}}{(2x+\alpha_{13}+1)(2y+\alpha_{24}+1)}\right\}
P_{x+1}^{(\alpha_3,\alpha_1-1)}(\cos 2\varphi_1)P_{y}^{(\alpha_4,\alpha_2-1)}(\cos 2\varphi_2)
\\
\qquad{}
\times
\Big([2x+\alpha_{13}+1]\frac{\sin^2\vartheta}{\cos^2\vartheta}P_{N-x-y}^{(2y+\alpha_{24}+1,2x+\alpha_{13}+1)}(\cos
2\vartheta)
\\
\qquad{}
+[N-x-y+1]\frac{1}{\cos^2\vartheta}P_{N-x-y+1}^{(2y+\alpha_{24},2x+\alpha_{13})}(\cos 2\vartheta)\Big)
\\
\qquad{}
+\left\{\frac{(x+\alpha_1)(y+1)(\cos^2\vartheta)^{x}(\sin^2\vartheta)^{y+1}}{(2x+\alpha_{13}+1)(2y+\alpha_{24}+1)}\right\}
P_{x}^{(\alpha_3,\alpha_1-1)}(\cos 2\varphi_1)P_{y+1}^{(\alpha_4,\alpha_2-1)}(\cos 2\varphi_2)
\\
\qquad{}
\times \Big([N-x-y+1]\frac{1}{\sin^2\vartheta}P_{N-x-y+1}^{(2y+\alpha_{24},2x+\alpha_{13})}(\cos 2\vartheta)
\\
\qquad{}
-[2y+\alpha_{24}+1]\frac{\cos^2\vartheta}{\sin^2\vartheta}P_{N-x-y}^{(2y+\alpha_{24}+1,2x+\alpha_{13}+1)
}(\cos 2\vartheta)\Big)
\\
\qquad{}
+\left\{\frac{(x+1)(y+1)(\cos^2\vartheta)^{x+1}(\sin^2\vartheta)^{y+1}}{(2x+\alpha_{13}+1)(2y+\alpha_{24}+1)}\right\}
P_{x+1}^{(\alpha_3,\alpha_1)}(\cos 2\varphi_1)P_{y+1}^{(\alpha_4,\alpha-2-1)}(\cos 2\varphi_2)
\\
\qquad{}
\times \Big([N-x-y+1]\frac{1}{\cos^2\vartheta\sin^2\vartheta}P_{N-x-y+1}^{(2y+\alpha_{24},2x+\alpha_{13})}(\cos
2\vartheta)
\\
\qquad{}
-[2y+\alpha_{24}+1]\frac{1}{\sin^2\vartheta}P_{N-x-y}^{(2y+\alpha_{24}+1,2x+\alpha_{13}+1)}(\cos 2\vartheta)
\\
\qquad{}
+[2x+\alpha_{13}+1]\frac{1}{\cos^2\vartheta}P_{N-x-y}^{(2y+\alpha_{24}+1,2x+\alpha_{13}+1)}(\cos 2\vartheta)\Big)\Bigg] .
\end{gather*}
The term between the parentheses are easily determined to be the following
\begin{gather*}
[2x+\alpha_{13}+1]\frac{\sin^2\vartheta}{\cos^2\vartheta}P_{N-x-y}^{(2y+\alpha_{24}+1,2x+\alpha_{13}+1)}(\cos
2\vartheta)
\\
\qquad
\phantom{=}{}
+[N-x-y+1]\frac{1}{\cos^2\vartheta}P_{N-x-y+1}^{(2y+\alpha_{24},2x+\alpha_{13})}(\cos 2\vartheta)
\\
\qquad{}
=(N-x+y+\alpha_{24}+1)P_{N-x-y}^{(2y+\alpha_{24},2x+\alpha_{13}+2)}(\cos 2\vartheta),
\\
[N-x-y+1]\frac{1}{\sin^2\vartheta}P_{N-x-y+1}^{(2y+\alpha_{24},2x+\alpha_{13})}(\cos 2\vartheta)
\\
\qquad
\phantom{=}{}
-[2y+\alpha_{24}+1]\frac{\cos^2\vartheta}{\sin^2\vartheta}P_{N-x-y}^{(2y+\alpha_{24}+1,2x+\alpha_{13}+1)
}(\cos 2\vartheta)
\\
\qquad{}
=-(N+x-y+\alpha_{13}+1)P_{N-x-y}^{(2y+\alpha_{24}+2,2x+\alpha_{13})}(\cos 2\vartheta),
\\
[N-x-y+1]\frac{1}{\cos^2\vartheta\sin^2\vartheta}P_{N-x-y+1}^{(2y+\alpha_{24},2x+\alpha_{13})}(\cos 2\vartheta)
\\
\qquad
\phantom{=}{}
-[2y+\alpha_{24}+1]\frac{1}{\sin^2\vartheta}P_{N-x-y}^{(2y+\alpha_{24}+1,2x+\alpha_{13}+1)}(\cos 2\vartheta)
\\
\qquad
\phantom{=}{}
+[2x+\alpha_{13}+1]\frac{1}{\cos^2\vartheta}P_{N-x-y}^{(2y+\alpha_{24}+1,2x+\alpha_{13}+1)}(\cos 2\vartheta)
\\
\qquad{}
=-(N+x+y+|\alpha|+3)P_{N-x-y-1}^{(2y+\alpha_{24}+2,2x+\alpha_{13}+2)}(\cos 2\vartheta).
\end{gather*}
Adjusting the normalization factors then yields the result~\eqref{Action-Ap}.

\subsection*{Acknowledgements}
V.X.G.\ holds an Alexander-Graham-Bell fellowship from the Natural Sciences and Engineering Research Council of Canada (NSERC).
The research of L.V.\ is supported in part by NSERC.

\pdfbookmark[1]{References}{ref}
\LastPageEnding


\begin{thebibliography}{99}
\footnotesize \itemsep=0pt

\bibitem{Alisauskas-2000}
Ali{\v{s}}auskas S., The triple sum formulas for {$9j$} coef\/f\/icients of {${\rm
  SU}(2)$} and {${\rm u}_q(2)$}, \href{http://dx.doi.org/10.1063/1.1312198}{\textit{J.~Math. Phys.}} \textbf{41} (2000),
  7589--7610, \href{http://arxiv.org/abs/math.QA/9912142}{math.QA/9912142}.

\bibitem{Alisauskas-1971}
Ali{\v{s}}auskas S.J., Jucys A.P., Weight lowering operators and the
  multiplicity-free isoscalar factors for the group {$R_{5}$}, \href{http://dx.doi.org/10.1063/1.1665626}{\textit{J.~Math.
  Phys.}} \textbf{12} (1971), 594--605.

\bibitem{Andrews_Askey_Roy_1999}
Andrews G.E., Askey R., Roy R., Special functions, \textit{Encyclopedia of
  Mathematics and its Applications}, Vol.~71, Cambridge University Press,
  Cambridge, 1999.

\bibitem{Arfken-2012}
Arfken G.B., Weber H.J., Mathematical methods for physicists, 5th ed.,
  Harcourt/Academic Press, Burlington, MA, 2001.

\bibitem{Bonzom-2012}
Bonzom V., Fleury P., Asymptotics of {W}igner {$3nj$}-symbols with small and
  large angular momenta: an elementary method, \href{http://dx.doi.org/10.1088/1751-8113/45/7/075202}{\textit{J.~Phys.~A: Math.
  Theor.}} \textbf{45} (2012), 075202, 20~pages, \href{http://arxiv.org/abs/1108.1569}{arXiv:1108.1569}.

\bibitem{Diaconis-02-2014}
Diaconis P., Grif\/f\/iths R., An introduction to multivariate {K}rawtchouk
  polynomials and their applications, \href{http://dx.doi.org/10.1016/j.jspi.2014.02.004}{\textit{J.~Statist. Plann. Inference}}
  \textbf{154} (2014), 39--53, \href{http://arxiv.org/abs/1309.0112}{arXiv:1309.0112}.

\bibitem{Dunkl-2001}
Dunkl C.F., Xu Y., Orthogonal polynomials of several variables,
  \href{http://dx.doi.org/10.1017/CBO9780511565717}{\textit{Encyclopedia of Mathematics and its Applications}}, Vol.~81, Cambridge
  University Press, Cambridge, 2001.

\bibitem{Genest-2013-06}
Genest V.X., Vinet L., Zhedanov A., The multivariate {K}rawtchouk polynomials
  as matrix elements of the rotation group representations on oscillator
  states, \href{http://dx.doi.org/10.1088/1751-8113/46/50/505203}{\textit{J.~Phys.~A: Math. Gen.}} \textbf{46} (2013), 505203, 24~pages,
  \href{http://arxiv.org/abs/1306.4256}{arXiv:1306.4256}.

\bibitem{Genest-2013-tmp-1}
Genest V.X., Vinet L., Zhedanov A., Superintegrability in two dimensions and
  the {R}acah--{W}ilson algebra, \href{http://dx.doi.org/10.1007/s11005-014-0697-y}{\textit{Lett. Math. Phys.}} \textbf{104}
  (2014), 931--952, \href{http://arxiv.org/abs/1307.5539}{arXiv:1307.5539}.

\bibitem{Zhedanov-09-1993}
Granovski{\u\i} Ya.I., Zhednov A.S., New construction of {$3nj$}-symbols,
  \href{http://dx.doi.org/10.1088/0305-4470/26/17/039}{\textit{J.~Phys.~A: Math. Gen.}} \textbf{26} (1993), 4339--4344.

\bibitem{Haggard-2010}
Haggard H.M., Littlejohn R.G., Asymptotics of the {W}igner {$9j$}-symbol,
  \href{http://dx.doi.org/10.1088/0264-9381/27/13/135010}{\textit{Classical Quantum Gravity}} \textbf{27} (2010), 135010, 17~pages,
  \href{http://arxiv.org/abs/0912.5384}{arXiv:0912.5384}.

\bibitem{Hoare-2008}
Hoare M.R., Rahman M., A probabilistic origin for a new class of bivariate
  polynomials, \href{http://dx.doi.org/10.3842/SIGMA.2008.089}{\textit{SIGMA}} \textbf{4} (2008), 089, 18~pages,
  \href{http://arxiv.org/abs/0812.3879}{arXiv:0812.3879}.

\bibitem{Kalnins-2006}
Kalnins E.G., Kress J.M., Miller Jr. W., Second order superintegrable systems
  in conformally f\/lat spaces. {IV}.~{T}he classical 3{D} {S}t\"ackel transform
  and 3{D} classif\/ication theory, \href{http://dx.doi.org/10.1063/1.2191789}{\textit{J.~Math. Phys.}} \textbf{47} (2006),
  043514, 26~pages.

\bibitem{Kalnins-2011-05}
Kalnins E.G., Miller Jr. W., Post S., Two-variable {W}ilson polynomials and the
  generic superintegrable system on the 3-sphere, \href{http://dx.doi.org/10.3842/SIGMA.2011.051}{\textit{SIGMA}} \textbf{7}
  (2011), 051, 26~pages, \href{http://arxiv.org/abs/1010.3032}{arXiv:1010.3032}.

\bibitem{Kalnins-1991}
Kalnins E.G., Miller Jr. W., Tratnik M.V., Families of orthogonal and
  biorthogonal polynomials on the {$N$}-sphere, \href{http://dx.doi.org/10.1137/0522017}{\textit{SIAM~J. Math. Anal.}}
  \textbf{22} (1991), 272--294.

\bibitem{Koekoek-2010}
Koekoek R., Lesky P.A., Swarttouw R.F., Hypergeometric orthogonal polynomials
  and their {$q$}-analogues, \href{http://dx.doi.org/10.1007/978-3-642-05014-5}{\textit{Springer Monographs in Mathematics}},
  Springer-Verlag, Berlin, 2010.

\bibitem{Koelink-1998}
Koelink H.T., Van~der Jeugt J., Convolutions for orthogonal polynomials from
  {L}ie and quantum algebra representations, \href{http://dx.doi.org/10.1137/S003614109630673X}{\textit{SIAM~J. Math. Anal.}}
  \textbf{29} (1998), 794--822, \href{http://arxiv.org/abs/q-alg/9607010}{q-alg/9607010}.

\bibitem{Lievens-2002}
Lievens S., Van~der Jeugt J., {$3nj$}-coef\/f\/icients of {${\rm su}(1,1)$} as
  connection coef\/f\/icients between orthogonal polynomials in {$n$} variables,
  \href{http://dx.doi.org/10.1063/1.1482149}{\textit{J.~Math. Phys.}} \textbf{43} (2002), 3824--3849.

\bibitem{VDJ-2003-2}
Lievens S., Van~der Jeugt J., Realizations of coupled vectors in the tensor
  product of representations of {$\mathfrak{su}(1,1)$} and
  {$\mathfrak{su}(2)$}, \href{http://dx.doi.org/10.1016/S0377-0427(03)00622-8}{\textit{J.~Comput. Appl. Math.}} \textbf{160} (2003), 191--208.

\bibitem{Miller-1968}
Miller Jr. W., Lie theory and special functions, \textit{Mathematics in Science
  and Engineering}, Vol.~43, Academic Press, New York~-- London, 1968.

\bibitem{Suslov-1991}
Nikiforov A.F., Suslov S.K., Uvarov V.B., Classical orthogonal polynomials of a
  discrete variable, \href{http://dx.doi.org/10.1007/978-3-642-74748-9}{\textit{Springer Series in Computational Physics}}, Springer-Verlag,
  Berlin, 1991.

\bibitem{Regge-2000}
Regge T., Williams R.M., Discrete structures in gravity, \href{http://dx.doi.org/10.1063/1.533333}{\textit{J.~Math.
  Phys.}} \textbf{41} (2000), 3964--3984, \href{http://arxiv.org/abs/gr-qc/0012035}{gr-qc/0012035}.

\bibitem{Rosengren-12-1998}
Rosengren H., On the triple sum formula for {W}igner {$9j$}-symbols,
  \href{http://dx.doi.org/10.1063/1.532634}{\textit{J.~Math. Phys.}} \textbf{39} (1998), 6730--6744.

\bibitem{Rosengren-1999}
Rosengren H., Another proof of the triple sum formula for {W}igner
  {$9j$}-symbols, \href{http://dx.doi.org/10.1063/1.533114}{\textit{J.~Math. Phys.}} \textbf{40} (1999), 6689--6691.

\bibitem{Rosengren-1998}
Rosengren H., Multivariable orthogonal polynomials and coupling coef\/f\/icients
  for discrete series representations, \href{http://dx.doi.org/10.1137/S003614109732568X}{\textit{SIAM~J. Math. Anal.}} \textbf{30}
  (1999), 232--272.

\bibitem{Z-2007}
Rudzikas Z., Theoretical atomic spectroscopy, \textit{Cambridge Monographs on
  Atomic Molecular and Chemical Physics}, Vol.~7, Cambridge University Press,
  Cambridge, 2007.

\bibitem{Rao-1989}
Srinivasa~Rao K., Rajeswari V., A note on the triple sum series for the {$9j$}
  coef\/f\/icient, \href{http://dx.doi.org/10.1063/1.528369}{\textit{J.~Math. Phys.}} \textbf{30} (1989), 1016--1017.

\bibitem{Nuc-2007}
Suhonen J., From nucleons to nucleus. Concepts of microscopic nuclear theory,
  \href{http://dx.doi.org/10.1007/978-3-540-48861-3}{\textit{Theoretical and Mathematical Physics}}, Springer, Berlin, 2007.

\bibitem{Tratnik-1991-04}
Tratnik M.V., Some multivariable orthogonal polynomials of the {A}skey
  tableau-discrete families, \href{http://dx.doi.org/10.1063/1.529158}{\textit{J.~Math. Phys.}} \textbf{32} (1991),
  2337--2342.

\bibitem{VDJ-1997}
Van~der Jeugt J., Coupling coef\/f\/icients for {L}ie algebra representations and
  addition formulas for special functions, \href{http://dx.doi.org/10.1063/1.531984}{\textit{J.~Math. Phys.}} \textbf{38}
  (1997), 2728--2740.

\bibitem{VDJ-02-2000}
Van~der Jeugt J., Hypergeometric series related to the {$9$}-{$j$} coef\/f\/icient
  of {${\mathfrak su}(1,1)$}, \href{http://dx.doi.org/10.1016/S0377-0427(00)00297-1}{\textit{J.~Comput. Appl. Math.}} \textbf{118}
  (2000), 337--351.

\bibitem{VDJ-2003}
Van~der Jeugt J., {$3nj$}-coef\/f\/icients and orthogonal polynomials of
  hypergeometric type, in Orthogonal Polynomials and Special Functions
  ({L}euven, 2002), \href{http://dx.doi.org/10.1007/3-540-44945-0_2}{\textit{Lecture Notes in Math.}}, Vol.~1817, Editors
  E.~Koelink, W.~Van~Assche, Springer, Berlin, 2003, 25--92.

\bibitem{Vilenkin-1991}
Vilenkin N.Ja., Klimyk A.U., Representation of {L}ie groups and special
  functions, \href{http://dx.doi.org/10.1007/978-94-017-2885-0}{\textit{Mathematics and its Applications}}, Vol.~316, Kluwer
  Academic Publishers Group, Dordrecht, 1995.

\bibitem{LittleJohn-2011}
Yu L., Littlejohn R.G., Semiclassical analysis of the Wigner $9j$ symbol with
  small and large angular momenta, \href{http://dx.doi.org/10.1103/PhysRevA.83.052114}{\textit{Phys. Rev.~A}} \textbf{83} (2011),
  052114, 14~pages, \href{http://arxiv.org/abs/1104.1499}{arXiv:1104.1499}.

\end{thebibliography}
\end{document}